\documentclass[12pt,notitlepage,a4paper]{article}
\pdfoutput=1
\usepackage{color,graphicx}
\usepackage{amssymb}
\usepackage{delarray,amsmath,bbm}
\usepackage[latin1]{inputenc}
\usepackage[american]{babel}
\usepackage{cite}
\usepackage{subfigure}
\usepackage{epsfig}

\newcommand{\be}{\begin{equation}}
\newcommand{\ee}{\end{equation}}
\newcommand{\beqa}{\begin{eqnarray}}
\newcommand{\eeqa}{\end{eqnarray}}
\numberwithin{equation}{section}

\newfont{\namefont}{cmr10}
\newfont{\addfont}{cmti7 scaled 1440}
\newfont{\boldmathfont}{cmbx10}
\newfont{\headfontb}{cmbx10 scaled 1728}

\begin{document}
\baselineskip=15.5pt
\pagestyle{plain}
\setcounter{page}{1}

\begin{center}
\rightline{IFT-UAM/CSIC-14-012 }
\vspace{0.4in}

\renewcommand{\thefootnote}{\fnsymbol{footnote}}

\begin{center}
{\Large \bf  Holographic Relaxation of  Finite Size Isolated Quantum Systems }
\end{center}
\vskip 0.1truein
\begin{center}
\bf{Javier Abajo-Arrastia${}^1$, Emilia da Silva${}^1$, 
\\ Esperanza Lopez${}^1$, 
Javier Mas${}^2$ and Alexandre Serantes${}^2$ }
\end{center}
\vspace{0.5mm}

\begin{center}\it{
${}^1$ Instituto de F\'\i sica Te\'orica IFT UAM/CSIC \\
C-XVI, Universidad Aut\'onoma de Madrid\\
 28049 Cantoblanco, 
 Madrid, Spain
}
\end{center}

\begin{center}\it{
${}^2$Departamento de  F\'\i sica de Part\'\i  culas \\
Universidade de Santiago de Compostela, 
and \\
Instituto Galego de F\'\i sica de Altas Enerx\'\i as IGFAE\\
E-15782 Santiago de Compostela, Spain}
\end{center}

\vspace{0.5mm}

\setcounter{footnote}{0}
\renewcommand{\thefootnote}{\arabic{footnote}}

\let\thefootnote\relax\footnotetext{Emails: javier.abajo@uam.es, emilia.dasilva.dit@gmail.com, esperanza.lopez@uam.es,\\
javier.mas@usc.es, alexandre.serantes@gmail.com}

\vspace{0.6in}

\begin{abstract}
\noindent 
We study holographically the out of equilibrium dynamics of a finite size closed quantum system in $2+1$ dimensions, modelled by the collapse of a shell of a massless scalar field in AdS$_4$. In global coordinates there exists a variety of evolutions towards final black hole formation which we relate with different patterns of relaxation in the dual field theory.   For large scalar initial data rapid thermalization is achieved as a priori expected. Interesting phenomena appear for small enough amplitudes. Such  shells do not generate a black hole by direct collapse, but quite generically,  an apparent horizon emerges after enough bounces off the AdS boundary. We relate this bulk evolution with relaxation processes at strong coupling which delay in reaching an ergodic stage. Besides the dynamics of bulk fields, we monitor the entanglement entropy, finding that it oscillates quasi-periodically before final equilibration. The radial position  of the travelling shell is brought in correspondence with the evolution of the pattern of entanglement  in the dual field theory. We propose, thereafter, that the observed oscillations are the dual counterpart of the quantum revivals studied in the literature.  The entanglement entropy is not only able to portrait the streaming of entangled excitations, but it is also a useful probe of interaction effects.

\end{abstract}

\smallskip
\end{center}

\newpage

\tableofcontents

\section{Introduction}
 Despite its fundamental importance, the relaxation  of closed quantum systems  is still  a  subject of debate both from the theoretical and the experimental perspective.
 The recent availability of highly controllable quantum simulators, together with the awareness of its conceptual importance, has stimulated the interest on 
  {\em quantum thermalization} (see \cite{Polkovnikov:2010yn} for a review with references).

To place this topic in a historical perspective, already in the classical realm,  interest in a similar question was behind the seminal  work of Fermi, Pasta and Ulam (FPU) on the dynamics of a one-dimensional anharmonic chain \cite{FermiPastaUlam}.
There, contrary to Fermi's expectation, the presence of nonlinearities was not enough to trigger the ergodicity required for a statistical behavior at long times. Fermi suspected  this was something deep and new, and indeed, this problem marks the starting point for two branches of classical dynamics that developed rapidly:  integrability and chaos. 

For quantum systems the situation is much less clear even at the theoretical level. From the experimental data, mounting evidence points towards a rich variety of evolutions, depending on the microscopic dynamics as well as on the initial conditions.
In some cases, like for hard core atomic interactions, integrability inhibits thermalization 
by freezing the momentum distribution, such that memory of the initial state is not lost  \cite{Kinoshita2006}.
In others, the system thermalizes after passing through  a quasi-stationary plateau at intermediate times, which has received the name of  prethermalization  \cite{Gring.et.al.2012}\cite{Trotzky.et.al.2012}.
Theoretical efforts have been put into trying to derive a statistical description for these quasi-stationary states by means of a generalised Gibbs ensemble    \cite{Jaynes1957}\cite{Rigol2007}\cite{Kollar2011}\cite{Smith2012}\cite{Mussardo2013}. Further work is still necessary to clarify the different routes from integrability to quantum chaos and quantum ergodicity. 
This paper aims to show that holographic techniques can contribute to the study of this fascinating subject.
 
The distinction between classical and quantum receives an  unexpected twist by means of the holographic duality. In short, it states that some quantum systems at strong coupling are believed to admit a description in terms of  a dual picture that involves classical General Relativity in one more dimension. 
Moreover, some non-local observables are computable from purely geometrical constructs. This is the case of the entanglement entropy, $S_A$, of a  region $A$ in space. It has been conjectured in \cite{Ryu:2006bv}, and recently  put on firmer grounds in\cite{Casini:2011kv} \cite{Lewkowycz:2013nqa}, to be given by the area of a minimal surface $\gamma_A$ which extends into the bulk while being homologous to $A$. More precisely
\be
S_A =\frac{ \hbox{Area} ( \gamma_A)}{4G_{N}} \, .
\label{ryutaka}
\ee
For non-static backgrounds the condition of minimal surface should be replaced by that of an extremal one \cite{Hubeny:2007xt}.

Entanglement entropy, which provides a measure of quantum entanglement in extended systems, is a notion receiving increasing attention. It has been studied in 
dynamical situations as a probe on the evolution towards equilibration.
The seminal work of Cardy and Calabrese  \cite{CardyCalabrese2007}\cite{Calabrese:2005in} focuses on quantum quenches from gapped to critical 1+1 field theories.
The authors prove that the entanglement entropy  grows with time until it saturates at  a constant value. The saturation time  increases proportionally  to the length of the interval, and the emerging picture is that of propagation of entanglement at the  speed of light.
Consistently, the same results have been recovered within a class of holographic models where the gravity dual involves a shell of null dust infalling from the AdS boundary  and forming a black hole deep inside \cite{AbajoArrastia:2010yt}\cite{Albash:2010mv}. This set up has been extended to higher dimensions \cite{Albash:2010mv}\cite{Balasubramanian:2010ce}\cite{Balasubramanian:2011ur} and to local quenches \cite{Nozaki:2013wia}. 

We want to push this venue further, and construct holographic models whose dynamics out of equilibrium  departs from a fast approach to ergodic behavior. We will focus on isolated quantum systems of finite size.
A simple example where neither equilibration nor thermalization  takes place involves  free fields with a linear dispersion relation on a circle. For such system any initial state will reconstruct periodically in time. In \cite{Takayanagi:2010wp} a dual counterpart of this behavior was proposed to be given by  a succession of quantum black holes that form and evaporate in an asymptotically global AdS space. 
Instead, a calculation involving only classical general relativity would be addressing  the problem of a strongly coupled quantum field theory living  in the boundary. 

Our central result is that an evolution pattern where the initial state is partially reconstructed several times before reaching equilibration, is also possible in strongly coupled theories. The simple gravitational system we will consider involves a massless scalar field coupled to Einstein gravity with negative cosmological constant.  
In \cite{Pretorius:2000yu}, working on AdS$_3$, collapse to a black hole was seen for amplitudes above a threshold in a similar fashion as found some years ago by Choptuik for asymptotically flat spacetimes  \cite{Choptuik1993}. A radical difference appears below  threshold. Now the compactness of global AdS, together with the fact that the scalar field has pressure (unlike the case of null dust), implies there should appear a periodic  regime where the scalar shell bounces  back and forth between the origin and the boundary. Althought this was indeed seen in \cite{Pretorius:2000yu}, a step further came out of the work by Piotr Bizo\'n and Andrzej Rostworowski on AdS$_4$ \cite{Bizon:2011gg}. They  pushed the simulations far enough in time and resolution so as to establish that, even for subcritical pulses,  after a large enough number of bounces, the solution ended up forming an apparent horizon.
We will propose to related this type of bulk solutions with a field theory dynamics able to retain quantum coherence on a long time scale.

The results of \cite{Bizon:2011gg} triggered the interest on this topic \cite{Jalmuzna:2011qw}-$\!\!$\cite{Bizon:2013xha}, and the subsequent analysis gave rise to a richer landscape.
Despite the initial suspect that horizon formation was the unavoidable end from evolving generic initial data, using perturbation theory, the authors of  \cite{Dias:2012tq} gave evidence  for the existence of  fully stable periodic non-linear solutions, and conjectured the existence of  ``islands of stability" around them. Some regular solutions were explicitly constructed  in \cite{Maliborski:2013jca} confirming the perturbative analysis. Also long lived, presumably stable, solutions were obtained in  \cite{Buchel:2013uba} (see also \cite{Maliborski:2013ula}).  In summary,  there is a rich landscape  of initial conditions, and it is very tempting to encompass this fact  with the variety of relaxation processes that are being observed in real closed quantum systems. It would be extremely interesting to setup a concrete dictionary.  This paper intends to be a step in that direction.

The nonlinear nature of the problem calls for the use of  numerical simulations.
 We have setup a  calculational  program that allows us to generate collapses of the above type and compute extremal surfaces on them. We have checked stability and convergence of our code, and  reproduced most of the results in the literature. We have focused on the AdS$_4$ case.  
Qualitatively,  higher dimensions share many relevant features \cite{Jalmuzna:2011qw}.  In fact  the physics is essentially the same as the one for the spherical scalar collapse on Minkowski space-time with reflective boundary conditions at some finite radius \cite{Maliborski:2012gx}. 
 
The paper is organised in the following way. Section 2 includes background material as well as a 
portrait of the landscape of collapse stories.  We have been able to prolong our simulations past the moment of horizon formation and, in some cases, until a black hole of almost the total scalar shell mass is stablished. Section 3 is devoted to a detailed dual interpretation of the bouncing solutions.
The space of initial conditions contains cases whose evolution looks either as a sharp packet propagation or as a radially delocalized wave. We relate these features
with the entanglement pattern in the dual field theory. 
In Section 4 we analyze the evolution of the entanglement entropy. The bouncing geometries induce an oscillatory pattern on it, which  follows from the quasi-periodic behavior 
of the metric. 
We show that the entanglement entropy can not only portrait kinematical effects, such as the streaming away of entangled excitations, but also interaction phenomena.
In section 5 we present our conclusions  and comment on the calculations that
are underway. 

\section{Scalar collapse}

We consider Einstein gravity with negative cosmological constant coupled to a real massless scalar field in four dimensions,
\be
S=\int d^{4}x \sqrt{g}\left( \frac{1}{16\pi G_4} R - 2\Lambda  - \frac{1}{2}\partial_\mu\phi \partial^\mu\phi\right)\, ,
\ee
with $\Lambda = -3/l^2$. We set $8\pi G_4 = 2$ as well as $l=1$. This system has been examined recently for the numerical study of gravitational collapse in asymptotically global AdS spaces, and we summarize some of the known results.
Our emphasis, however, is set in pushing the simulation beyond the apparent horizon formation. With the aim at motivating our interpretation in the dual theory, we pursue the evolution as far as the numerical code permits us to approach the final stationary black hole. 

\subsection{Equations of motion}

We will follow the ansatz and conventions in \cite{Bizon:2011gg} for a spherically symmetric collapse. For the line element this gives
\be
ds^2 = \frac{1}{\cos^2 x}\left( - A(t,x) e^{-2\delta(t,x)} dt^2 + {dx^2 \over A(t,x)} + \sin^2 x\, d\Omega_{2}^2\right) \, ,
\label{metric}
\ee
where  $x\in[0,\pi/2]$ is a compact radial coordinate, and $d\Omega_{2}^2$ stands for the metric  of the unit sphere. The equations of motion can be casted in a first order form
\beqa
\dot\Phi &=& \left( A e^{-\delta} \Pi\right)' ~~~,~~~\dot \Pi = \frac{1}{\tan^{2} x}\left(\tan^{2} x\, A e^{-\delta} \Phi\right)' \, ,\label{eqforphi}
\\
A' &=&\displaystyle \frac{1 + 2 \sin^2 x}{\sin x \cos x} (1-A) -\sin x \cos x \, A\,  (\Phi^2 + \Pi^2)\, ,\label{eqforA}\\[2mm]
\delta' &=& -\sin x \cos x \,(\Phi^2 + \Pi^2)\, . \label{eqford}
\eeqa
where $\Phi\!=\!\phi'$ and $\Pi\!=\!A^{-1} e^\delta \dot{\phi}$, with $\phi'$ and  $\dot{\phi}$ the space and time derivatives of the scalar field respectively.  Equations \eqref{eqforphi} are  evolution equations for the scalar field, and   \eqref{eqforA} is the Hamiltonian constraint. Due to  isotropy, there are no evolution equations for the metric components. 

We want to solve the previous system of differential equations given smooth  initial data for the scalar field. Regularity at the origin implies  $A(t,0)\!=\!1$, which fixes the integration constant from \eqref{eqforA}.
Equation \eqref{eqford} is invariant under time dependent shifts of the function $\delta(t,x)$. This is equivalent to reparameterizing the time direction, a gauge freedom left over by the ansatz \eqref{metric}. Since we are interested in the holographic dictionary, a natural way to fix this freedom is by requiring $t$ to be the proper time at the boundary,
namely $\delta (t,\pi/2)\!=\!0$. 
Further imposing that the total mass remains constant along the evolution sets to zero the non-normalizable mode of the scalar, associated with a source term in the dual field theory lagrangian\footnote{The scalar field can also take a constant value at infinity of no physical relevance.}. Under these conditions, the expansion of the fields close to the origin is
\be
\phi = \phi_0(t)+{\cal O}(x^2) \, , \hspace {5mm} A=1+{\cal O}(x^2) \, , \hspace {5mm} \delta = \delta_0(t) +{\cal O}(x^2) \, ,
\label{exorigin}
\ee
whereas close to the boundary one finds
\be
\phi = \phi_\infty(t) y^3 + {\cal O}(y^{5}) \, , \hspace {5mm} A=1 - 2M y^3 + {\cal O}(y^{6}) \, , \hspace {5mm} \delta = {\cal O}(y^{6}) \, ,
\label{exboundary}
\ee
with $y\!=\!\pi/2\! -\! x$. The total mass is
\be
M = {1 \over 2} \int_0^{\pi/2} dx \, \rho(t,x)   \, , \hspace{1cm}  \rho(t,x)=\tan^{2}\! x \, (\Phi^2 + \Pi^2)  \, e^{-\delta}\, ,
\label{mass}
\ee
where $\rho$ provides a description of the radial energy distribution of the scalar pulse.\footnote{The integrand of \eqref{mass} is not uniquely defined. The choice ${\tilde \rho}= \tan^{2}\! x \, (\Phi^2 + \Pi^2)  \, A$ reconstructs equally $M$ \cite{Bizon:2011gg}. Both functions give quite similar results.}  

Concerning initial conditions, they will be set as values for $\Pi(0,x)$ and $\Phi(0,x)$.
The numerical integration of equations \eqref{eqforphi}-\eqref{eqford} subject to the above described initial and boundary conditions is accomplished using a finite difference method  with a fourth order Runge-Kutta algorithm. In addition to the 
equations \eqref{eqforphi}\eqref{eqforA} and \eqref{eqford} there is an additional ``momentum constraint"
$\dot A + 2\sin x\cos x A^2 e^{-\delta} \Phi \Pi=0$. Fulfillment of this equation and constancy of the mass will be used as quality check of our numerical simulations.

The choice of the slice $t\!=\!0$ as initial data surface is suited to the observable we want to study, namely the entanglement entropy. Ideally, we would require that its holographic derivation at any positive boundary time does not require information from across our initial data surface. As recalled in \eqref{ryutaka}, the entanglement entropy is captured by the area  of extremal  surfaces  in the bulk that anchor on the boundary of AdS to the boundary of a chosen region. In a static spacetime, these extremal surfaces are minimal surfaces and can be shown to live on constant $t$ slices. Although this does not hold for a dynamical background,  deviations are small in the cases we consider here as we will see below. For comparison, an alternative choice is to set initial data on the null infalling surface $v=0$, where $v$ is an Eddington-Finkelstein coordinate (see for example  \cite{Heller:2011ju}\cite{Heller:2012km}). This would be inconvenient in our case, as extremal surfaces ending at a boundary time $t\!\gtrsim\!0$ would pierce that surface towards negative values of $v$.
 
Instead of performing a scan over the full space of initial conditions for the scalar field, we have restricted to  gaussian type profiles localized either close to the origin of AdS
\be
\Phi_c(x)=0\, , \hspace{1cm} \Pi_c(x)={2 \epsilon \over \pi}  \exp\left( {\displaystyle-{4 \tan^2 x \over \pi^2 \sigma^2}} \right)\, .
\label{profileO}
\ee
as in \cite{Bizon:2011gg}, or close to the boundary
\be
\Phi_b (x)=0\, , \hspace{1cm} \Pi_b (x)={12 \epsilon \over \pi}  \exp\left({-{4 \tan^2 (\pi/2 - x) \over \pi^2 \sigma^2}}\right) \cos^3 x\, .
\label{profileB}
\ee
These second type of boundary conditions, although exhibiting a very similar phenomenology to the first type, looks more akin to 
the Vaidya setup that has been used to model an analog to a quench in the dual field theory. 

Altogether, our initial conditions are therefore parameterised by two variables, the amplitude $\epsilon$ and the width $\sigma$. Of course, the space of initial conditions is infinite dimensional, and may hide
surprises that deserve to be unveiled. Some of them involve stationary regular solutions with and without rotation \cite{Dias:2012tq} and, from them, at least one has been  constructed numerically \cite{Maliborski:2013jca}.

\subsection{Collapse portrait}

Black holes formed in the collapse of a spherical shell of a massless real scalar field are of  Schwarzschild type and  can have either positive or negative specific heat. With our conventions, the threshold mass separating both cases is \footnote{Schwarzschild-AdS$_4$ black holes have a temperature
$T\!=\!(3 \tan^2 x_h\!+\!1)/ \tan x_h$, where $x_h$ is the largest root of $2 M \cos^2 x\!=\!\tan x$. The threshold mass is set by requiring $dT/dM\!=\!0$.} 
\be 
M_{th}=  {2 \over 3 \sqrt{3}} \sim 0.385\, .
\label{mthres}
\ee
Big (small) black holes refer to those with masses above (below) this threshold, which correspondingly have positive (negative) specific heat.
Due to the negative specific heat, small black holes can not be in thermal equilibrium with their own Hawking radiation, and they will evaporate very much as they do in flat space. For small $G_N$ however, 
Hawking radiation is suppressed and the process of evaporation becomes very slow as compared with the  time scales we will be interested in. Moreover this type of configurations has been shown to have higher entropy than thermal AdS with the same energy \cite{oai:arXiv.org:1105.4167}. Hence in the microcanonical ensemble we are considering, both small and large AdS black are valid collapse end products.

Regardless of which one of the  profiles \eqref{profileO} or \eqref{profileB} are used as initial data, for masses above or comparable to $M_{th}$, the direct formation of a black hole of the total mass is observed. Decreasing the mass neatly below $M_{th}$, the emerging apparent horizon starts  trapping only a fraction of the scalar  pulse, while the rest scatters towards the boundary. Upon lowering the amplitude further, a critical black hole (actually a naked singularity) of vanishing radius (hence mass) can form. This is in total agreement with the findings by Choptuik   \cite{Choptuik1993} in asymptotically flat space.  

The presence of an apparent horizon is signalled by a zero of the function $A(t,x)$.  However this function does not strictly vanish at any finite value of $t$, since as it drops the relative redshift factor with respect to the boundary grows very large  and the dynamics gets frozen around the emergent horizon. Hence in the coordinate system we are using, it only makes sense to define the time of horizon formation  as that when the minimum of $A(t,x)$ drops below a sufficiently small value. In this sense it should be understood below.

\begin{figure}[h]
\begin{center}
\includegraphics[width=7.5cm]{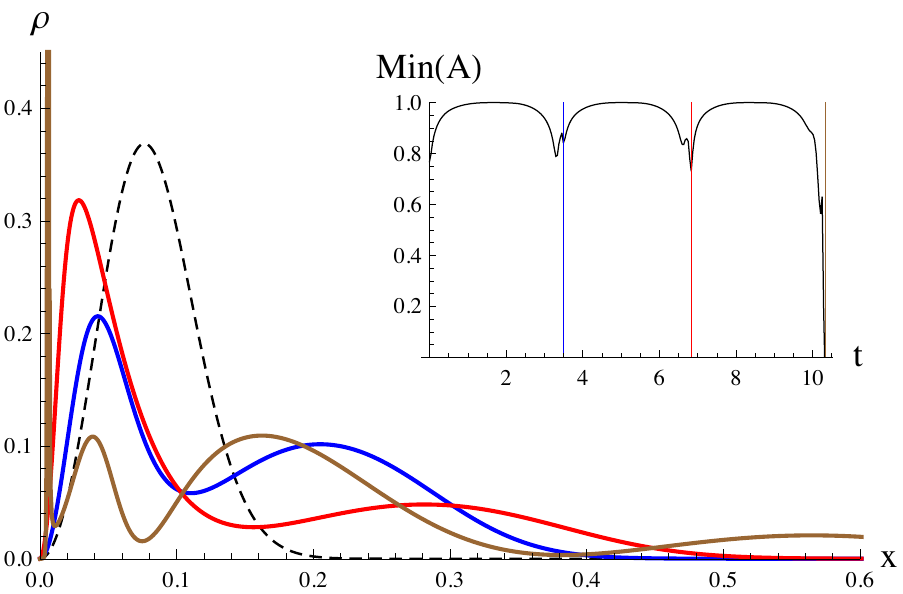}~
\includegraphics[width=7.5cm]{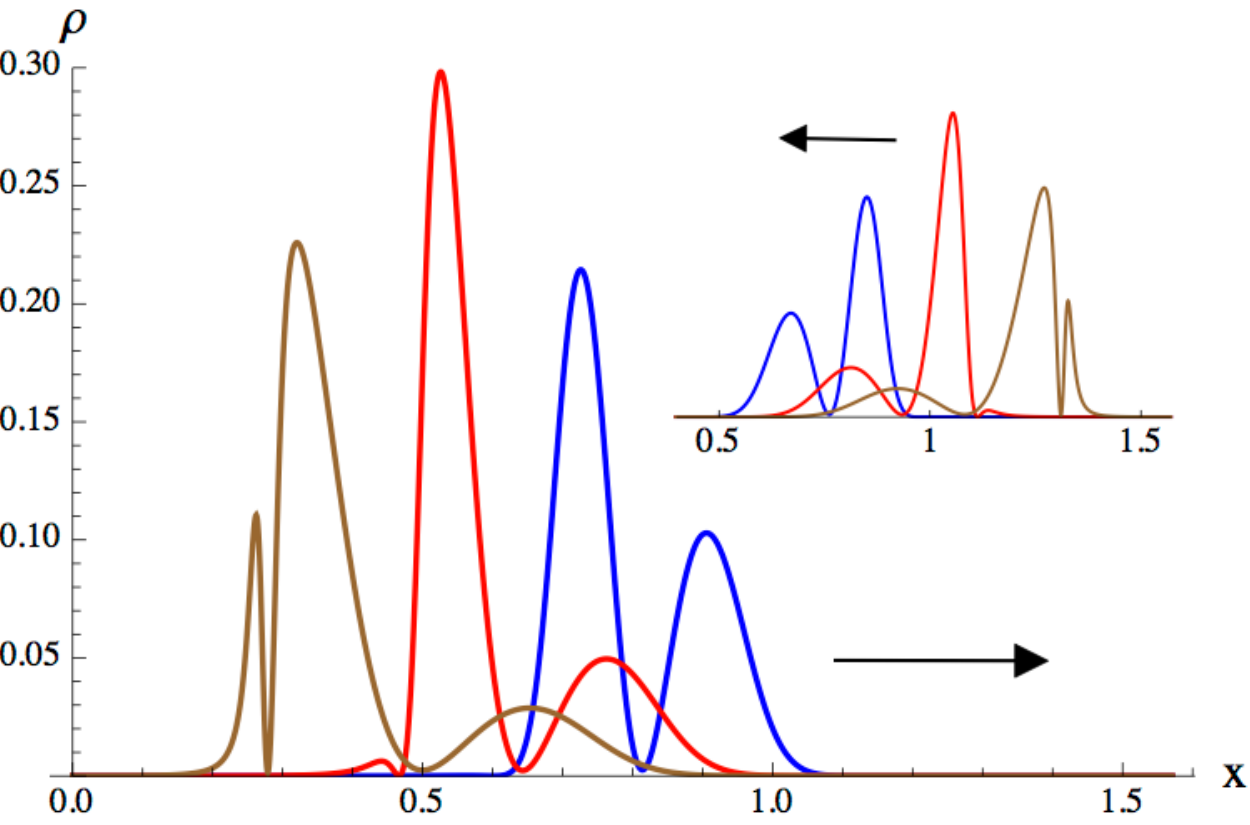}
\end{center}
\caption{\label{fig:AdSins} Evolution of a narrow pulse \eqref{profileO} with $\sigma\!=\!1/16$ and $M\!=\!0.015$. Left: the dashed line shows the initial mass distribution function. The curves in color denote the mass density profile at the times when the pulse bounces against the origin producing a minimum of $A(t,x)$ (see inset). Field values grow wild with time at the origin, enhancing  the non-linear effects. Right: scalar profiles from the first (blue), second (red) and third (brown) bouncing cycles; the arrows indicate the direction of movement. 
The scalar profile changes its shape when scattering at the origin, while it travels almost unaltered along the full radial coordinate.}
\end{figure}

New effects appear below the threshold amplitude for critical collapse. In flat space, all the mass would  just be reflected back to infinity and that would be the end of the story. Here instead, the massless scalar pulse that escapes towards the boundary of $AdS$ reflects in finite proper time, and falls in again.  In \cite{Bizon:2011gg}\cite{Jalmuzna:2011qw} support for a remarkable phenomenon was provided for AdS$_{d+1}$ when $d\geq 3$: well localized scalar profiles of arbitrary small initial amplitude always generate a horizon after a sufficient number of bouncing cycles. The mechanism behind this phenomenon is the transfer of energy from long to small wavelengths along the evolution \cite{Bizon:2011gg}\cite{Dias:2011ss}. Analogous effects are well known from fluid dynamics \cite{Yudovich:1974}, where they are referred to as weak turbulence.
This fact is responsible for the change of shape of the traveling wave and the sharpening of at least one of its fronts. Its action is most effective at the origin, where the bouncing produces extremely high values of the fields, therefore enhancing the effect of the nonlinearities.
After enough number of bounces a sufficiently sharp subpulse freezes, and the radial minimum of $A(t,x)$ drops abruptly (see inset in Fig.\ref{fig:AdSins}a). This is the defining time for an apparent horizon formation, although as said before, in these coordinates the exact formation occurs in infinite time. 

We would like to point out an effect accompanying  the transfer of energy towards small wavelengths. 
Each time the signal scatters at the origin and a part of it sharpens, the rest instead tends to increase its radial dispersion.
This behavior, which will prove to have important consequences for the dual field theory, is illustrated in Fig.\ref{fig:AdSins}a using a narrow initial profile which requires three bounces for collapse. There we have plotted the mass distribution function $\rho(t,x)$ \eqref{mass} at the times of closest approach to the origin. On Fig.\ref{fig:AdSins}b we illustrate how the scalar profile changes shape when it scatters at the origin, while its shape remains largely unaffected upon propagating and reflecting against the boundary. The profiles in color blue, red and brown correspond respectively to configurations along the first, second and third bouncing cycles.
The apparent horizon is generated from the small spiky front in the brown profile.

\begin{figure}[h]
\begin{center}
\includegraphics[width=7.8cm]{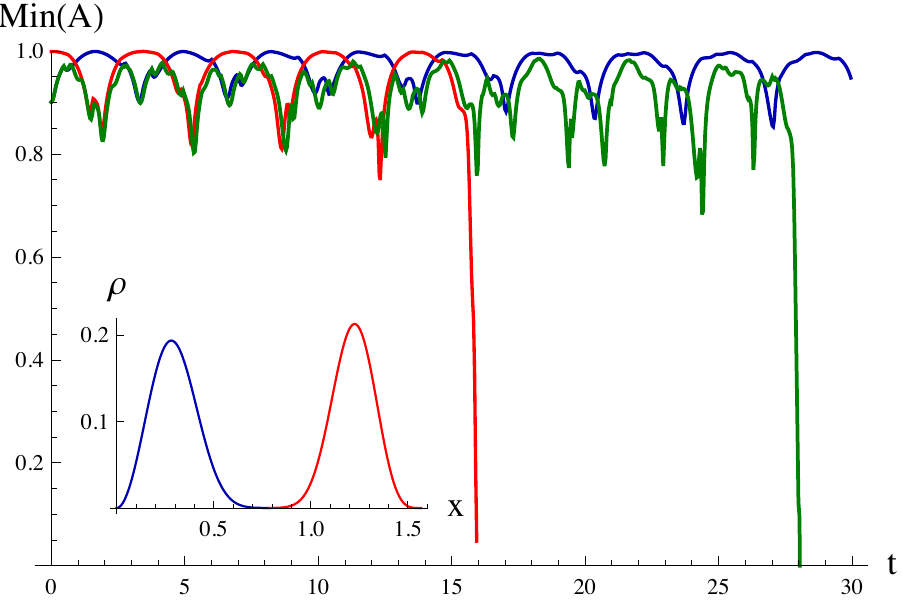} 
\hspace{5mm}
\includegraphics[width=7.8cm]{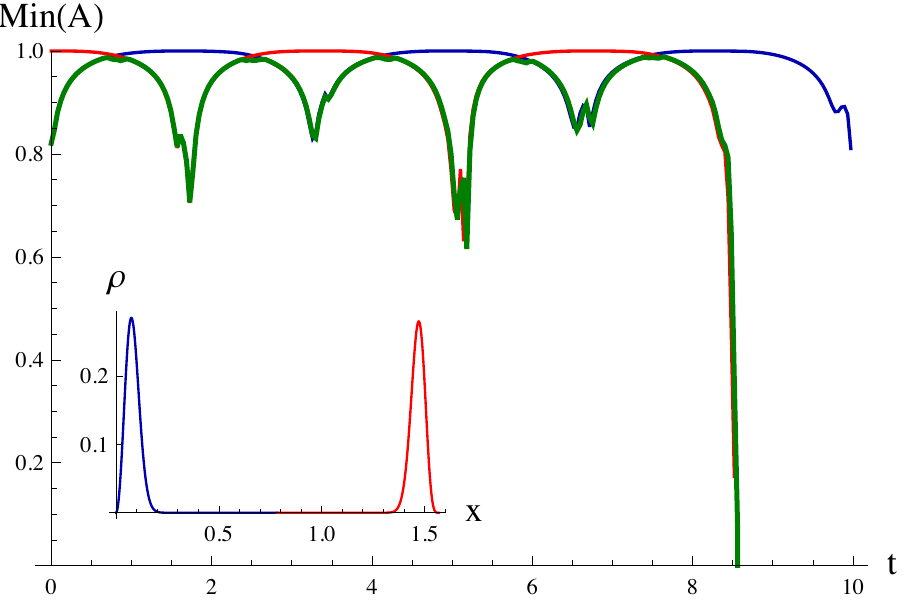}
\end{center}
\caption{\label{fig:int} Evolution of the initial data \eqref{profileO} (blue), \eqref{profileB} (red) and the combined profile \eqref{profileOB} (green) for the following subpulse data. Left: $\sigma\!=\!0.25$ and $M\!=\!0.012$. Right: $\sigma\!=\!1/16$ and $M\!=\!0.029$. In the inset, the initial mass density function. The left initial data have more overlap than the right ones. The horizon formation time for the fast collapsing pulse (red curve) gets affected and delayed (green curve), whereas it remains unaltered in the non-overlapping case on the right.}
\end{figure}

Besides the mass, a parameter which has strong influence on the evolution of a scalar pulse is its broadness. The relevance of this parameter was made manifest in \cite{Buchel:2013uba}. It was shown that the turbulent mechanism characteristic of the narrow pulse evolution becomes less efficient as the broadness of the profile grows.  Broad pulses quickly develop a subpulse structure with infalling and outgoing components which scatter among themselves. In order to better understand the interaction among subpulses, let us consider a different initial profile: a linear combination of \eqref{profileO} and \eqref{profileB} producing two well localized pulses close to the origin and the boundary 
\be
\Phi(x)=0\, , \hspace{1cm} \Pi(x)=\Pi_c (x)+\Pi_b (x)\, .
\label{profileOB}
\ee
We choose them to have the same broadness and the same mass. When the tails of the initial pulses have some overlap, the creation of an apparent horizon is delayed with respect to the independent evolution of the pulse that would collapse first,  $\Pi_b$. We observe in Fig.\ref{fig:int}a that the number of bounces necessary for the creation of a horizon increases. Hence scattering tends to work against weak turbulence, which in this example finally wins. Interestingly, pulses with a negligible initial superposition are practically transparent to each other, as shown in   Fig.\ref{fig:int}b. 

\begin{figure}[h]
\begin{center}
\includegraphics[scale=0.6]{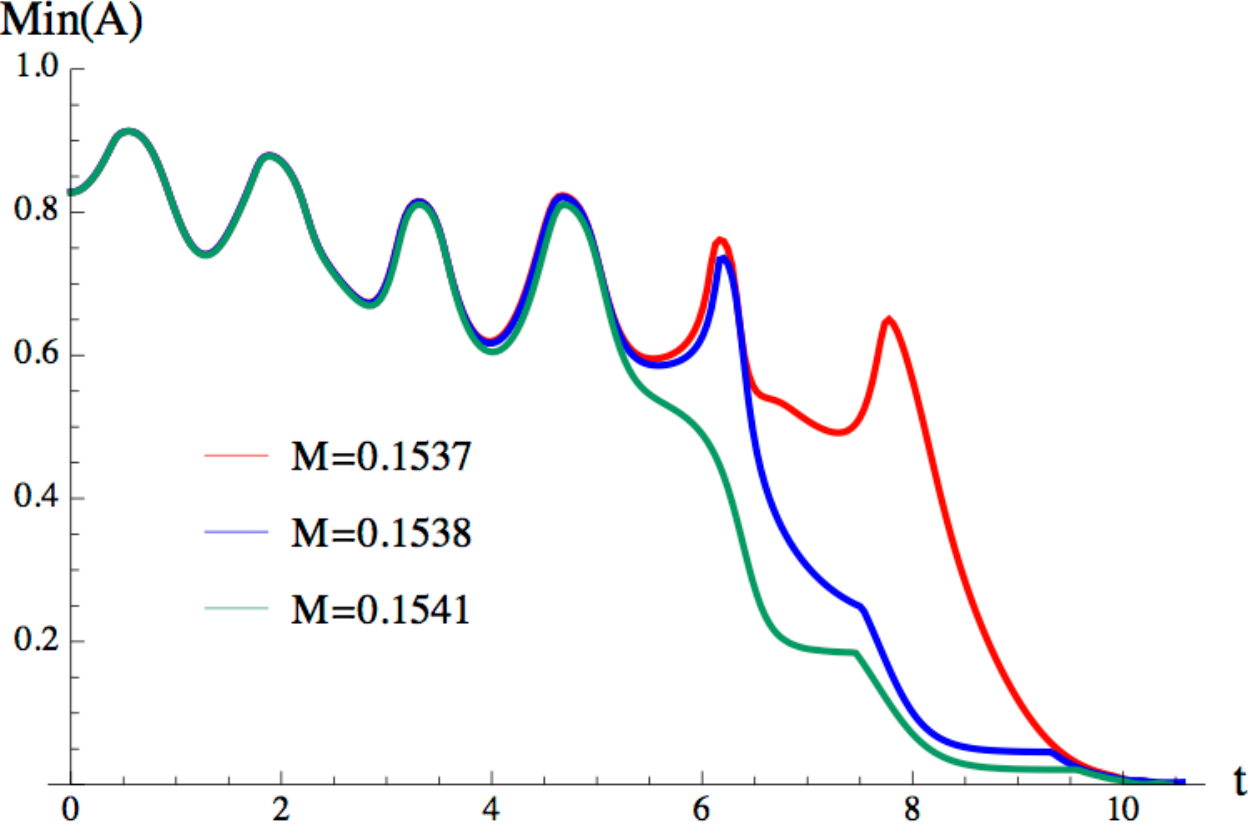} ~~
\includegraphics[scale=0.62]{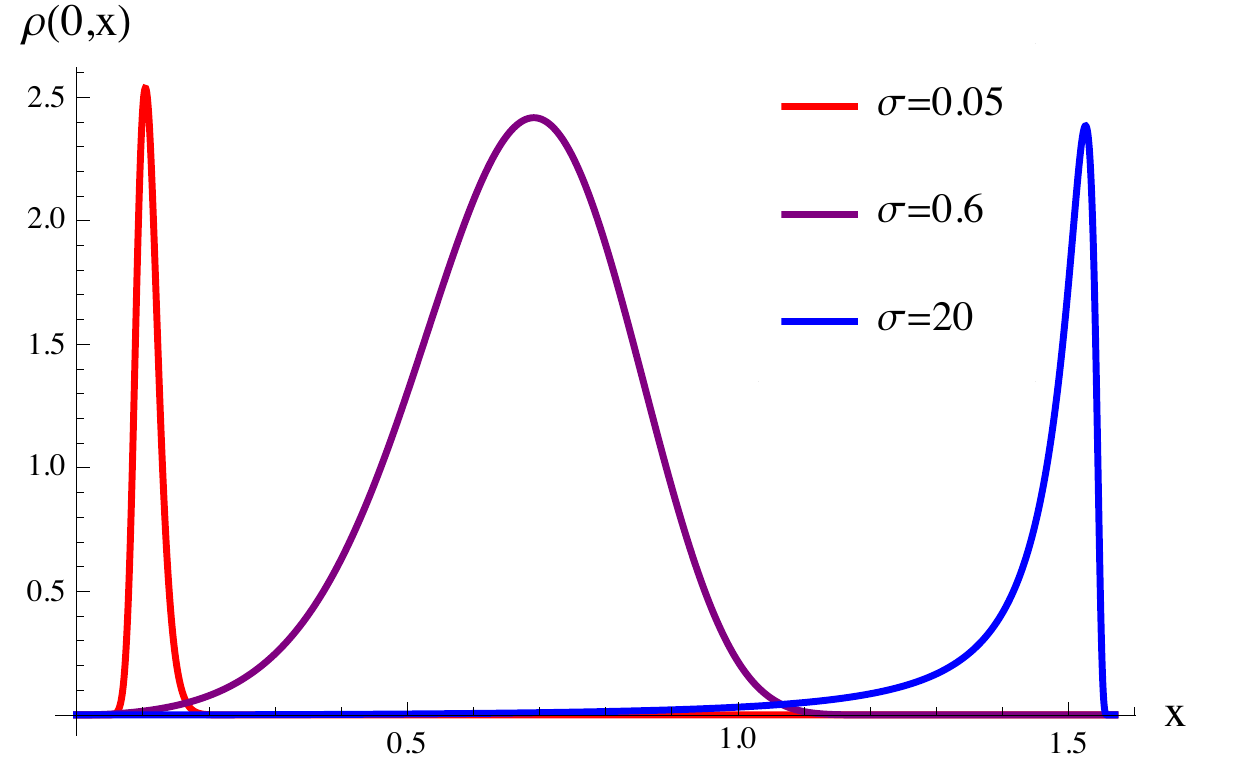}
\end{center}
\caption{\label{fig:sharpvsbroad} Left: Broad initial profiles of the form  \eqref{profileO} with $\sigma\!=\!0.6$. Right: influence of the parameter $\sigma$ on the initial mass density function $\rho(0,x)$. $\epsilon$ has been adjusted to make all curves similar in height.}
\end{figure}

Collapse processes with very broad initial profiles present distinguished characteristics. They are delocalized along the complete radial direction in the major part of the evolution. 
The oscillation periodicities of these solutions are determined, besides radial displacement, by their internal subpulse structure. As a result, the bouncing cycles are not neatly defined, see Fig.\ref{fig:sharpvsbroad}a. Moreover, the horizon emerges supported by a finite fraction of the pulse mass, in contrast to narrow pulses where it can be vanishingly small.
Delocalized pulses require masses around $40\%$ the threshold mass \eqref{mthres} to generate an apparent horizon. When the total mass is decreased, a point is reached where the time elapsed until horizon formation abruptly increases \cite{Buchel:2013uba}\cite{Maliborski:2013ula}. 
For masses below this threshold our results join previous analysis supporting the establishment of a regular quasi-standing wave \cite{Dias:2012tq}-$\!\!$\cite{Maliborski:2013ula}.

For initial conditions of the form \eqref{profileO} the threshold broadness for the existence of regular solutions appear to be around $\sigma\!\sim\!0.35$ \cite{Buchel:2013uba}.
In \cite{Maliborski:2013ula} the numerical analysis at $\sigma=0.6$ confirmed the regularity of the evolution. However, for large values of 
$\sigma$, again collapse to black hole formation is observed.  Their interpretation thereafter is that, for some intermediate values of $\sigma$, the evolution lies within the domain of attraction of  stable periodic solution that bifurcate from the fundamental mode of the linearized scalar wave equation in AdS, which the authors in  \cite{Buchel:2012uh} termed  ``oscillons''. For the lowest frequency in AdS$_4$ they are given by
\be
\phi^{osc}_{j=0}(t,x) =  a \cos(3 t + \alpha) \cos^3 x\, , \label{osci}
\ee
with  constants $a$ and $\alpha$. 
For $\alpha\!=\!\pi/2$, the initial conditions
\be
\Pi^{osc}_{j=0}(0,x) =3 a \cos^3 x \, , \hspace{8mm} \Phi^{osc}_{j=0} =0\, ,
\ee
are actually very similar to those coming from \eqref{profileO} with $\sigma\!\sim \!0.6$. 

On the other hand, the sharpness or broadness of the initial profile should be established on the mass density function, $\rho(x)$, rather than on the profile of the scalar field. We observe from the Fig.\ref{fig:sharpvsbroad}b that indeed,  sharp localized profiles exist both at small and at large values of $\sigma$.
Similar reasoning can be applied  to initial conditions of the form \eqref{profileB}. 
When $\sigma\!\gtrsim\!2$ the exponential factor in that profile can be neglected, and 
$\Pi_b(0,x) \propto \Pi_{j=0}^{osc} (0,x)$.
We have checked that for small enough amplitudes the evolution of these initial data is regular up to the reach of our computational capabilities.

\subsection{Post-horizon evolution}

Previous numerical simulations have been stopped at the moment when an apparent horizon forms. We need however to pursue the evolution as close as possible to the set up of the final static black hole in order to give a dual interpretation to the collapse processes we are studying. Although the dynamics around an emergent horizon gets practically frozen, the part of the scalar pulse that escapes its trapping effect continues traveling and reflecting towards the boundary. We have been able to describe this dynamics using the same coordinate system $(t,x)$ by increasing the resolution of the radial grid. Using grids with up to $10^5$ points we have obtained post-horizon evolutions within an acceptable precision. Namely we have checked that the momentum constraint  remains under control, the total mass of the configuration keeps constant up to few percent, and all function converge smoothly under variations of the resolution. 

\begin{figure}[h]
\begin{center}
\includegraphics[width=8cm]{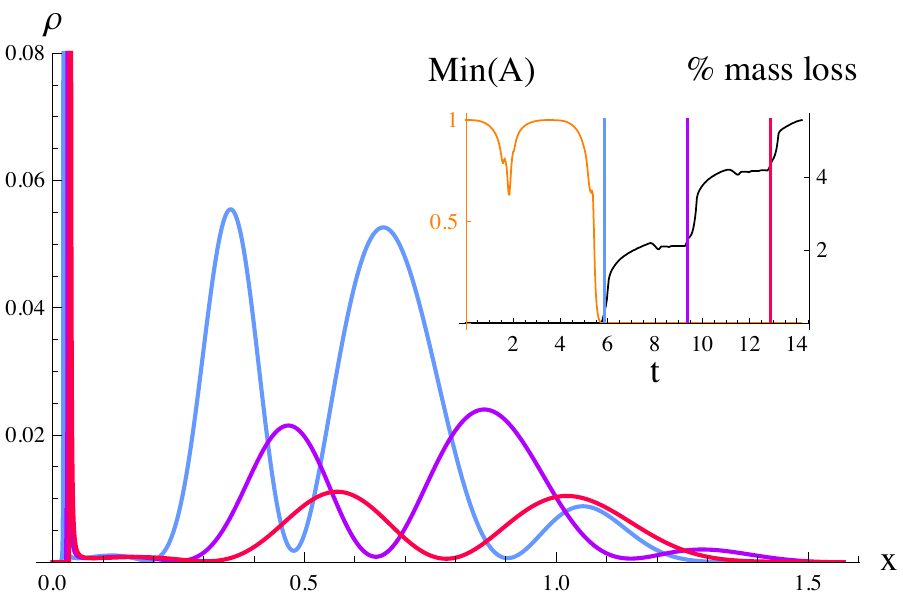} 
\includegraphics[width=8cm]{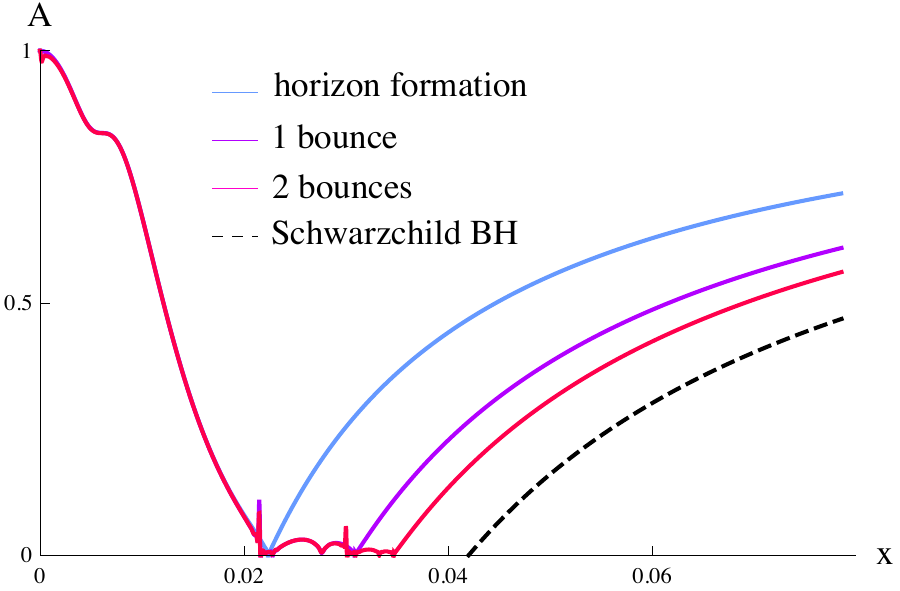}
\end{center}
\caption{\label{fig:AdSpost} Evolution of a narrow pulse \eqref{profileB} with $\sigma\!=\!0.1$ and $M\!=\!0.021$. Left: leftover scalar pulse after horizon formation and two post-collapse bounces. Right: horizon growth.}
\end{figure}

In Fig.\ref{fig:AdSpost} we plot the post-horizon evolution of a narrow profile  for which $A(t,x)$ abruptly drops after one bounce (see inset). The horizon radius at that moment is approximately half the one of a Schwarzschild BH of the total mass. When the apparent horizon emerges, the leftover scalar profile has already started its way to the AdS boundary. The mass distribution function at this moment is shown in Fig.\ref{fig:AdSpost}a (light blue). Subsequently this outgoing fraction of the pulse bounces with the boundary and falls in again, being   partially absorbed and partially reflected. As a result of this, a new minimum of $A(t,x)$ drops to zero at a larger value of $x$, in accordance with the expected growth of the horizon. This  is shown in Fig.\ref{fig:AdSpost}b. We managed to follow this process past the completion of a second bounce with an acceptable precision. The total mass of the scalar pulse, obtained by integrating $\rho(t,x)$ and which should keep constant along the evolution, only suffers a $2\%$ loss in each post-collapse cycle for a grid of $5\times 10^4$ points (see inset). The numerical noise around the minima of $A$ in Fig.\ref{fig:AdSpost}b could be linked to the small mass loss.

The horizon radius after two bounces is still far from that of a Schwarzschild BH of the total mass (dashed black line in Fig.\ref{fig:AdSpost}b). Several further bouncing cycles appear to be necessary to complete the collapse process. This is likely a generic feature. At a linearized level one can calculate the absorption of scalar field to be consistent with an outside amplitude that decreases exponentially with time $ |\phi|_{out} \sim \exp{(- \omega_l t)}$ with $\omega_l \sim r_h^2$, \cite{Horowitz:1999jd}\cite{Berti:2009wx}. Of course, deriving this dependence in the full non linear setup is an interesting question  we hope to report on in the near future.\footnote{We would like to thank Vitor Cardoso for clarification on this point.}

As it was the case for the pre-horizon dynamics, the post-horizon counterpart also depends on the broadness of the scalar profiles.
Pulses of intermediate broadness show some degree of localization along the pre-horizon evolution. 
However once an apparent horizon emerges, the fraction of the scalar pulse left out looses radial localization and a damped quasi-standing wave sets in.
This is indeed consistent with a result presented in the previous subsection. Namely, the weak turbulence mechanism implies, together with the progressive sharpening of a fraction of the scalar pulse, the radial dispersion of the rest.

\begin{figure}[h]
\begin{center}
\includegraphics[width=7.8cm]{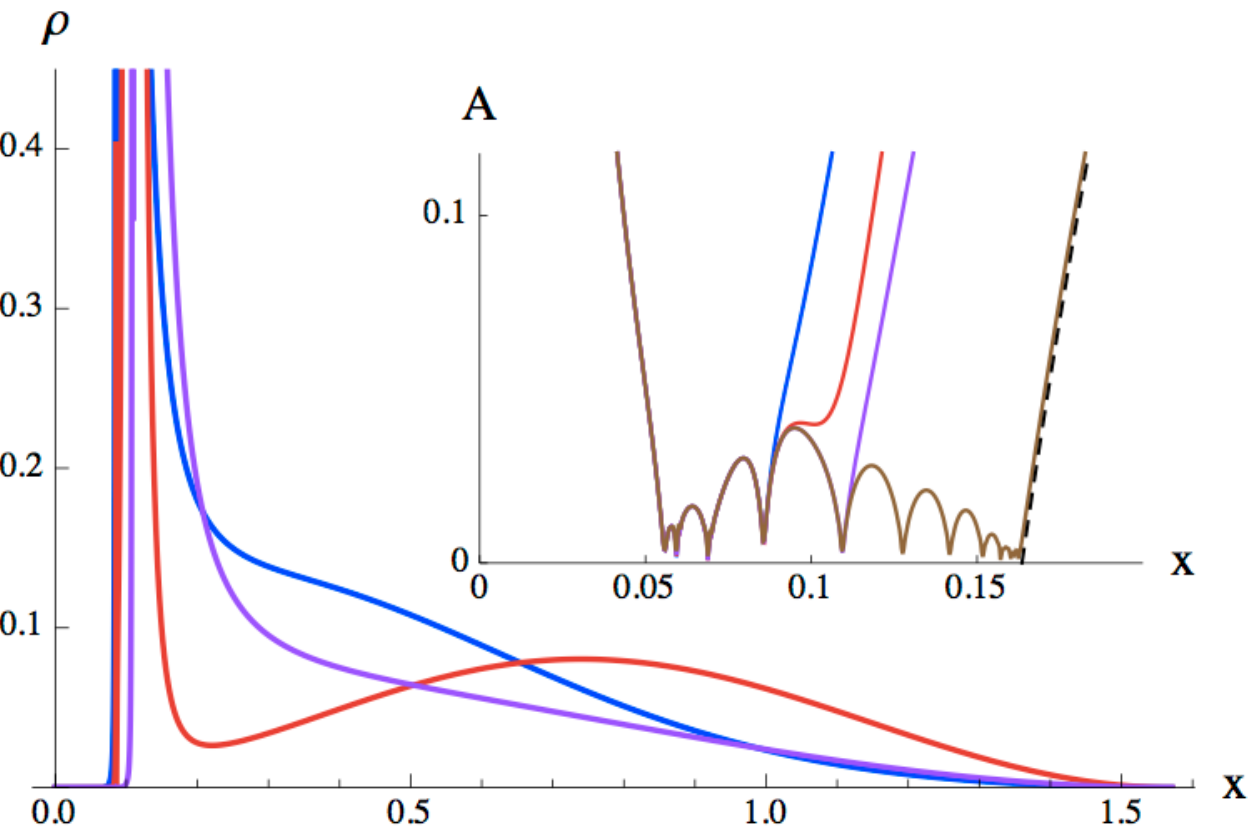} 
\hspace{5mm}
\includegraphics[width=7.8cm]{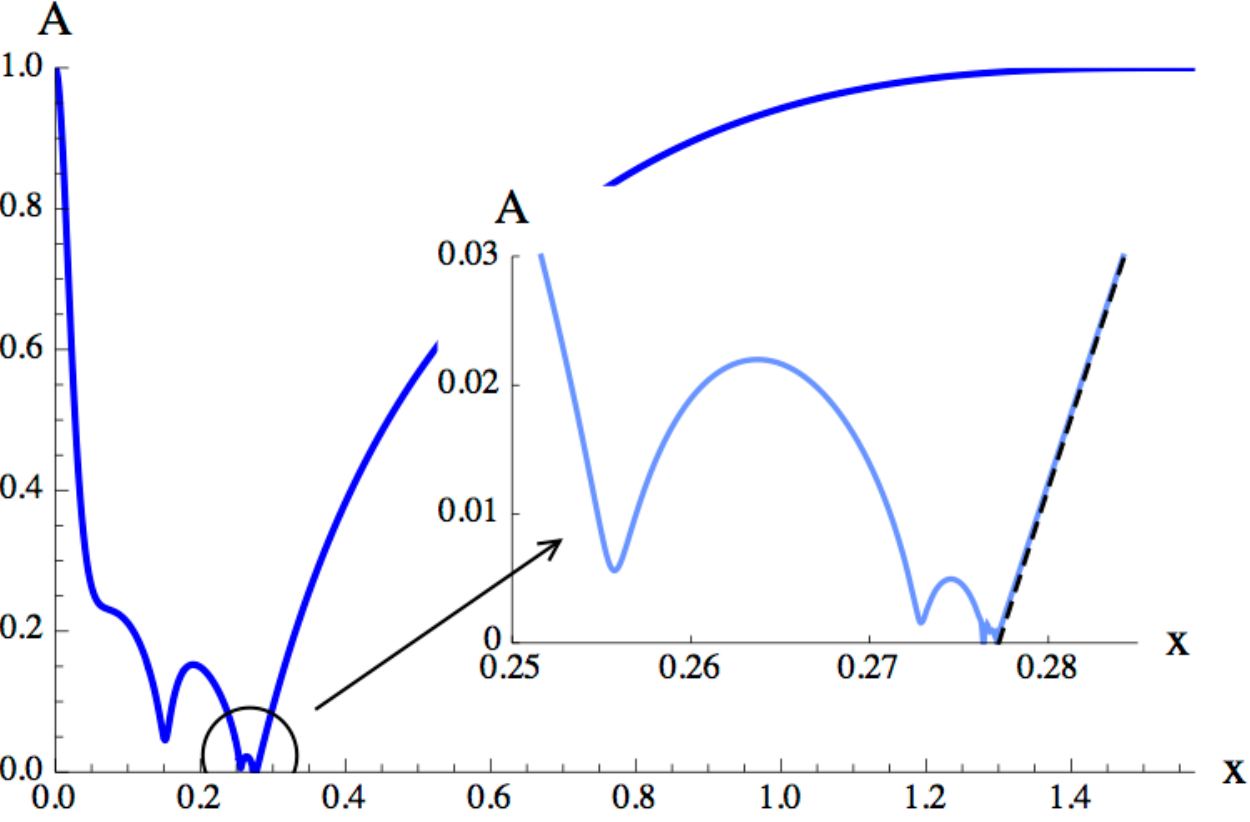}
\end{center}
\caption{\label{fig:medium} Left: cycle of the post-horizon damped wave for a profile \eqref{profileO} with $\sigma\!=\!0.25$ and $M\!=\!0.085$. In the inset we have plotted $A(t,x)$ 
when the collapse process is nearly completed. The dashed black line shows $A(x)$ for a Schwarzschild BH of the total mass. Right: same plot as in the inset for a profile \eqref{profileO} with $\sigma\!=\!0.6$ and $M\!=\!0.1538$.}
\end{figure}

An example of this behavior is shown in Fig.\ref{fig:medium}a for an initial profile \eqref{profileO} with $\sigma\!=\!0.25$.
After two well defined bouncing cycles, the value of $A(t,x)$ suddenly drops. As the evolution continues new minima of $A$ appear at growing values of the radial coordinate, until the final value given by the radius of a Schwarzschild black hole of the total mass is reached (see inset). We have plotted a complete cycle of the left-out  quasi-standing wave (blue, red and magenta curves), which perfectly matches the generation of new $A$ minimum. 
Following the post-horizon evolution is less demanding numerically for broad that for narrow pulses. We could reach a horizon radius up to within $3\%$ of  the final value in our example while keeping the mass loss around $0.1\%$ using a grid with $2\times 10^4$ points.

In the case of radially delocalized pulses, the function $A(t,x)$ does not drop abruptly to zero. This can be observed in Fig.\ref{fig:sharpvsbroad}a for several pulses with 
$\sigma=0.6$. A dynamics similar to the quasi-standing wave of Fig.\ref{fig:medium}a, but with a stronger damping, is stablished when the minimal radial value of $A$ stops oscillating and starts decreasing. However, the minimum of $A$ only becomes vanishingly small at a radial position very close to the final horizon radius. This can be appreciated in Fig.\ref{fig:medium}b for the scalar pulse with $M\!=\!0.1538$ from Fig.\ref{fig:sharpvsbroad}a.

The periodicity of the bouncing cycles is a very important information for the dual interpretation of the collapse processes.
The bouncing period of narrow pulses is always bigger but close to $\pi$, increasing with the mass and broadness of the scalar profile\footnote{The bouncing period is practically $\pi$  when measured with respect to proper  time at the origin\cite{Bizon:2011gg}. However for a boundary observer more energetic pulses take a longer time to climb up their own gravitational potential.}.
Instead, when the pulses are broad and dynamics is radially delocalized, a shorter 
periodicity emerges. It governs the metric oscillations of the broad initial profiles shown in Fig.\ref{fig:sharpvsbroad}a, as well as the damped quasi-stationary wave of Fig.\ref{fig:medium}. 
The presence of a faster oscillation should be traced back to the internal dynamics of the delocalized scalar pulse, rather than to radial propagation. The value of its period is always bigger but not far from $\pi/3$. 
The origin of this  frequency is related to the large overlap that  broad profiles have with the regular periodic solution found in \cite{Maliborski:2013jca}, which, in turn, branches out from the lowest linear ``oscillon''  \eqref{osci}. The periodicity of  \eqref{osci} is actually $2\pi /3$. However if we consider the backreaction of this scalar configuration on the metric, which is sourced  by the squares of the field derivatives, its resulting natural period is $\pi /3$.

\begin{figure}[h]
\begin{center}
\includegraphics[width=7.5cm]{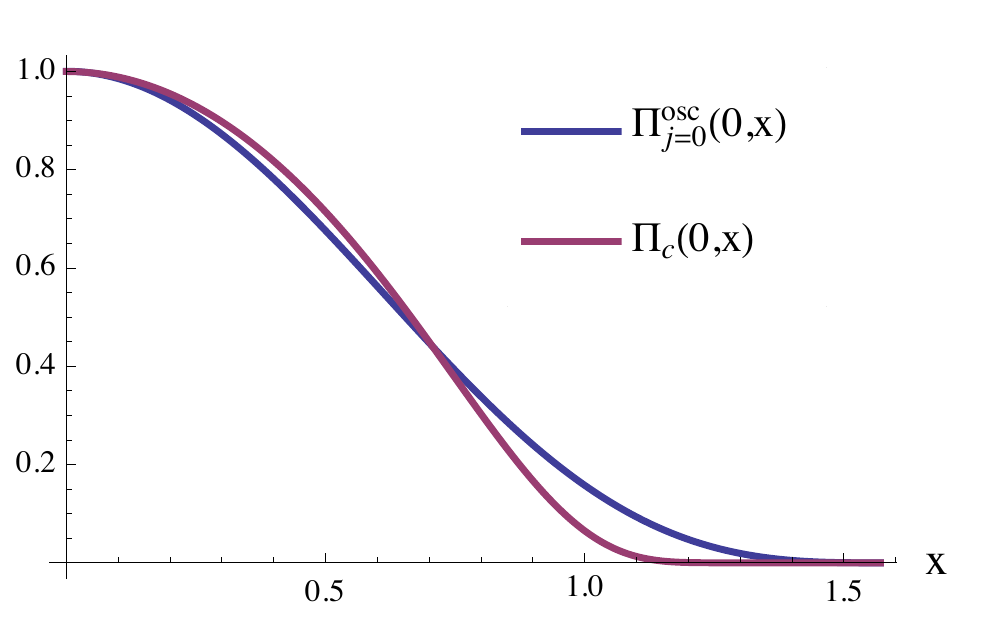} 
\end{center}
\caption{\label{fig:overlap} Initial data for $\Pi_c(0,x)$ with $\sigma = 0.6$ and $\Pi^{osc}_{j=0}(0,x)$ for the lowest  linear ``oscillon'' given in  \eqref{osci}, chosen to have the same height. The overlap is substantial, supporting the argument that their posterior evolutions, for small amplitudes, are in the same island of stability around the (close to) $w_0 = 3$ periodic  nonlinear solution constructed in \cite{Maliborski:2013jca}. }
\end{figure}
The appearance of this frequency is mysterious from the point of view of the quantum field theory. As a matter of fact, a hint on some experimental quantum observable which may have frequencies in the ``oscillon''  sequence, $\omega_j = d+2j$, would have
breathtaking implications  pointing towards a more than accidental relevance of the AdS/CFT correspondence.

\section{Dual interpretation of the bounces}

In order to propose a field theory interpretation for the bouncing geometries treated in the previous section,
let us first recall the holographic model of thermalization based on a Vaidya metric. Vaidya metrics describe the collapse of a shell composed of null dust. We consider now Poincar\'e instead of global coordinates, such that the dual field theory lives on Minkowski space. For simplicity we focus on the AdS$_3$/CFT$_2$ instance of the duality, and the limit in which the infalling shell is infinitely thin. The resulting geometry is that of a BTZ black hole outside the shell and empty AdS inside.
This model describes a sudden action on the dual CFT vacuum which creates a homogeneous plasma with non-trivial quantum correlations 
and its subsequent unitary evolution \cite{AbajoArrastia:2010yt}. It has been used as a holographic analog of a quantum quench.

\begin{figure}[h]
\begin{center}
\includegraphics[width=13cm]{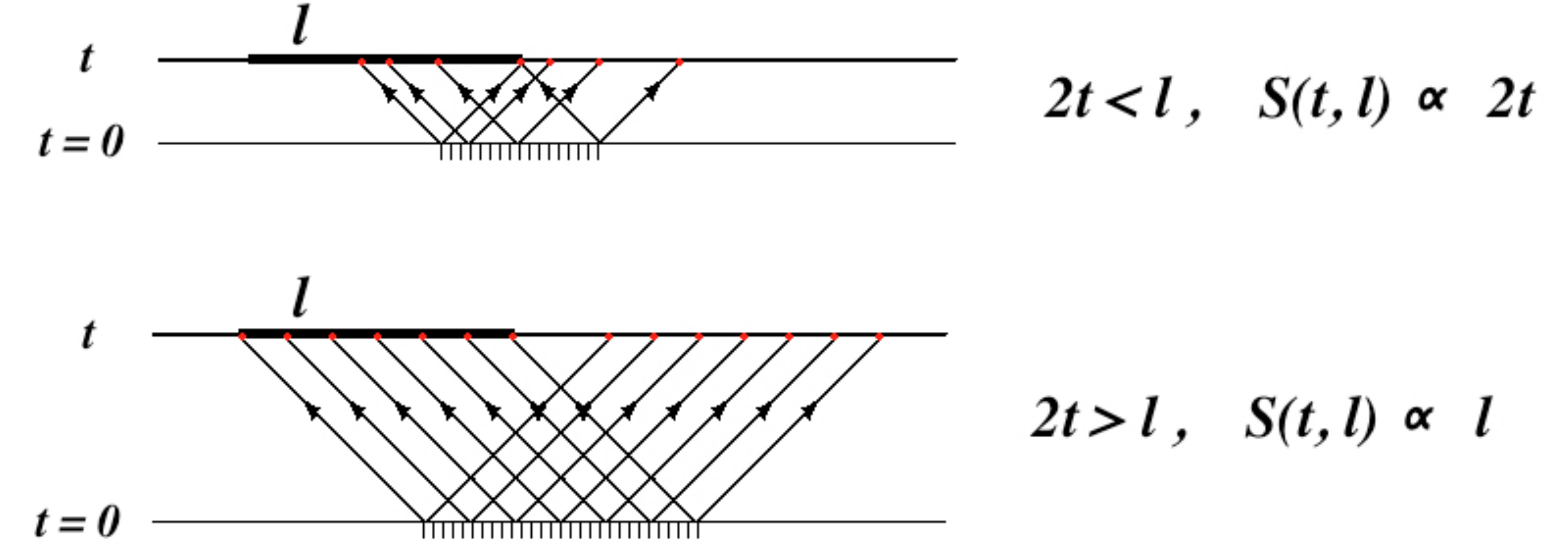} 
\end{center}
\caption{\label{fig:CC} Assuming that the excitations behave as free quasiparticles, the EE of an interval, $S(t,l)$, is proportional to the number of entangled pairs (shaded region) such that one component lies on the chosen interval and the other on its complementary \cite{Calabrese:2005in}. For simplicity it is considered that only excitations created at the same point are entangled.}
\end{figure}

In 2005 P. Calabrese and J. Cardy  studied quantum quenches from a gapped to a critical  system in 1+1 dimensions \cite{Calabrese:2005in}.
They showed that the entanglement entropy (EE) of a single interval of size $l$ increases with time until it saturates around
\be
t=l/2 \, .
\label{lth}
\ee
The limiting value attained at later times scales extensively precisely as it would do if  the system inside the interval was in contact with  a thermal bath given by the complementary  system outside.
Remarkably this behavior can be explained kinematically,  as a mere result of entangled excitations flying apart  at the speed of light (see Fig.\ref{fig:CC}).

Consistently, the holographic model based on the collapse of a null dust shell leads to the same saturation time \eqref{lth} for the EE as found in the CFT computation.
Adapting from \eqref{ryutaka}, the entanglement entropy of an interval 
is given by the length of the bulk geodesic that anchors at the AdS boundary on the interval endpoints. 
For small enough intervals or late enough times, geodesics will lie outside the infalling shell. Since this part of the geometry is that of a BTZ black hole, the entanglement entropy for such intervals will be the same as that of a thermal state. The bound \eqref{lth} corresponds to geodesics whose central point just reaches the infalling shell (either because $l$ is large enough, or because $t$ is sufficiently small) \cite{AbajoArrastia:2010yt}\cite{Balasubramanian:2010ce}. 
Similar results apply to the extremal surfaces computing the EE of spherical regions in 2+1 and 3+1 dimensions. They reach the infalling shell at
\be
t=R\, ,
\ee 
where $R$ is the radius of the sphere \cite{Albash:2010mv}\cite{Balasubramanian:2010ce}\cite{Balasubramanian:2011ur}.
We are thus led to propose a meaning for the position of the pulse in the radial direction as capturing the 
typical separation of entangled components  of the  QFT wavefunction.
When the pulse is close to the boundary,
entanglement is to be stronger between nearest neighbours,  
whereas the pulse falling towards the origin of AdS should represent entangled excitations flying apart. 
Analogous arguments were used in \cite{Nozaki:2013wia} to construct a holographic model for a local quench. This interpretation departs from the standard
lore that would tend to think of the radial position as encoding the characteristic energy of excitations in the perturbed state.

There is an important difference between the gapped-to-critical quenches studied in \cite{Calabrese:2005in} and the holographic Vaidya model.  In the former case, quantum correlations in the initial out of equilibrium state are short range, of the order of the inverse mass gap before the quench. 
The Vaidya model represents an homogeneous action on the CFT vacuum which creates a plasma with long range correlations. Actually the entanglement entropy and two-point functions just after a sudden perturbation practically coincide with that of the CFT vacuum  \cite{AbajoArrastia:2010yt}\cite{Balasubramanian:2010ce}.  This implies that correlations are strongest among excitations sourced at neighbouring points and decay with their distance as a power law. 
Hence the minimal separation of entangled components coincides with the distance across which quantum correlations are stronger. 
It is this distance that we relate with the radial position of the shell. The long range correlations are instead encoded in the AdS geometry interior to the shell.

\subsection{Dephasing and self-reconstruction}

An important notion in out of equilibrium physics is that of dephasing time. This is the time that a system takes to lose quantum coherence. Since we are dealing with closed systems, this notion will rather refer to the moment at which entanglement becomes inaccessible to macroscopical observables.
After dephasing, the system is expected to relax to a stationary state, generically a thermal state. 

The linear growth in threshold time \eqref{lth} implies that no matter how long we wait after a quench in an infinite system, there are always large enough regions where quantum coherent correlations can be detected. Namely, dephasing is never achieved at the global level. With the aim at providing a dual interpretation for the bouncing geometries studied in the previous section,
we review now the different scenarios that can arise on a compact space.
As motivated above, we assume that the initial state that triggers the field theory evolution presents stronger entanglement among neighbouring degrees of freedom. The typical separation of entangled components will start growing much as it does in the non-compact case. What happens after the maximal separation is achieved depends however, crucially,  on  interactions.

The simplest case to consider is that of a non-interacting theory with linear dispersion relation living on a circle. Any initial state reconstructs itself periodically, preventing the system to equilibrate.
When the initial state is homogeneous, this periodicity is 
\be
t_0={L \over 2v}\, 
\label{period}
\ee
with $L$ the length of the circle and $v$ the propagation velocity  of the excitations \cite{Takayanagi:2010wp} \cite{Cazalilla2006}. We will refer to $t_0$ as propagation time, since it is the time that  two particles moving apart with speed $v$ on the circle take to meet again.
 
The pattern in an interacting field theory is expected to differ substantially. As entangled excitations created at close points reach maximal separation on the circle, and naively start approaching again, interactions will have generically randomized their relative phases preventing the initial state from reconstructing \cite{Takayanagi:2010wp}. A 
strongly interacting holographic version of this behavior is provided by the 3d collapse of a thin spherical shell of null dust.  
The absence of pressure induces the formation of a black hole by direct collapse, and the EE of half the circle achieves a final value at 
\be 
t={L \over 4}\, .
\ee
This time consistently equals the one needed by two particles that separate at the speed of light to reach opposite points on the circle.

The evolution of dephasing not only hinges upon the microscopic dynamics, but also  depends upon the structure of the  initial state. Let us illustrate this statement with an easy example, by looking at the behavior of free systems with a non-linear dispersion relation. This is the case of a periodic chain of coupled harmonic oscillators 
\be
H={1 \over 2} \sum_{i=1}^N \Big[ \pi_i^2 + \nu^2 (\phi_{i+1}-\phi_i)^2 \Big]\, .
\label{harmonic}
\ee
Its spectrum is given by non-interacting modes of momentum $p\!=\!2 \pi n / N$ with $n\!=\!0,\dots,N\!-\!1$ and frequency
\be
\omega_p=2\nu \sin {p \over 2}  \, .
\ee
For $p\!\ll\!\pi$ the dispersion relation becomes linear, $\omega_p\!\simeq\!\nu p$. An initial wave packet constructed out of low momenta will reconstruct itself with period $t_0\!=\!N/2\nu$, as in \eqref{period}. No sign of relaxation will appear until enough time has passed to render important the non-linearity of the dispersion relation. If the wave packet is centered around frequency $\bar \omega$, this time is \cite{Polkovnikov:2010yn}
\be
t \simeq  {N^2 \over {\bar \omega}}\, .
\ee
Afterwards the system dephases and tends to a stationary state.\footnote{The stationary state generically differs from thermal equilibrium since the occupation numbers of non-interacting modes are conserved along the evolution} The dephasing time can be much larger than the propagation time $t_0$ if $\bar \omega$ is chosen sufficiently small.
It is important to emphasize here that this dependence of the relaxation process  on the initial state is indeed seen in experimental setups  \cite{Kinoshita2006} \cite{Gring.et.al.2012}\cite{Trotzky.et.al.2012}\cite{Kollar2011}.

We can now state the main proposal of this work: {\em  collapses  which require bouncing on the AdS boundary before forming a horizon are holographically dual to field theory evolutions where the initial state is partially reconstructed several times before achieving equilibration}.

In other words, the dephasing time is much larger than the typical propagation time, in analogy with the case of the harmonic chain above. We have argued that in an interacting field theory this is not the generic behavior.
However, that reasoning can fail for states with small enough energy density. The finite size of the system introduces an intrinsic scale and, hence,  the dynamical process can also depend on the energy density of the initial state. This is precisely what is found in the holographic models based on the collapse of a massless scalar profile \cite{Bizon:2011gg}. When the mass of the scalar shell is above the threshold \eqref{mthres} for the formation of a large black hole, the shell is completely trapped behind a horizon by direct collapse. Bouncing with the AdS boundary is only required when the final black hole to be formed is small. 

In the same way that an infalling scalar pulse is to be holographically interpreted as a growing separation between entangled excitations, the stages of the evolution when it moves towards the AdS boundary should represent entangled excitations joining again.
This can be neatly seen using a thin shell of null dust that travels outwards and reaches the boundary at $t\!=\!0$. The same reasoning that for an infalling shell sets the bound \eqref{lth}, leads now to
\be
0\leq l\leq-2t \, ,
\ee
for the size of intervals producing a thermal result for the entanglement entropy. Hence their size decreases as the system evolves towards $t\!=\!0$, as can be predicted from  the qualitative picture in Fig.\ref{fig:CC}.

A very nontrivial support for this picture comes from the periodicity of the scalar pulse in the bulk. From the numerical simulations, one can see that its evolution from the boundary to the center and back  completes a full  roundtrip with a period of approximately $\pi$ (see for example the inset in figure \ref{fig:AdSins}a). Now, recalling that in the gravitational system we have fixed the radius of the boundary sphere to unity, the expected reconstruction time \eqref{period} is 
\be
t_0={L \over 2}=\pi
\ee
where $L = 2\pi$ is the length of an equator. 
As we have mentioned in Section 2, the exact periodicity is slightly bigger than $\pi$ and this delay increases with the amplitude of the pulse. 
We will relate this fact to the presence of interactions in Section 4.1.

Besides dephasing, another process which is crucial in order to achieve thermal equilibrium is the equipartition of energy among degrees of freedom. 
If the radial position of the traveling shell would also describe this process, we should expect an oscillatory pattern in the evolution of the occupation numbers. 
However periodicities in occupation numbers are more naturally associated with Poincare recurrences, and in a large $N$ system these take a much longer time than $L/2$.
Hence we are led to conclude that the radial displacement of the scalar shell is not directly related with the redistribution of energy, and its periodic behaviour is tantamount to 
the reconstruction, or  revival,  of the initial dual quantum state. Evidence for revivals has been observed experimentally in cold atoms systems \cite{Greiner2002}. At a theoretical  level, it has been seen in quantum spin  chains \cite{Igloi2011}\cite{Happola2012} and also in   CFT in 1+1 dimensions \cite{Cardy2014}.

\subsection{Broadness versus time span} 

In our set up no reference is made as to how the initial state that triggers the evolution was created. In order to understand the implications of broad scalar profiles, it is however useful to discuss rough characteristics of field theory perturbations that can holographically relate to them.

\begin{figure}[h]
\begin{center}
\includegraphics[width=7cm]{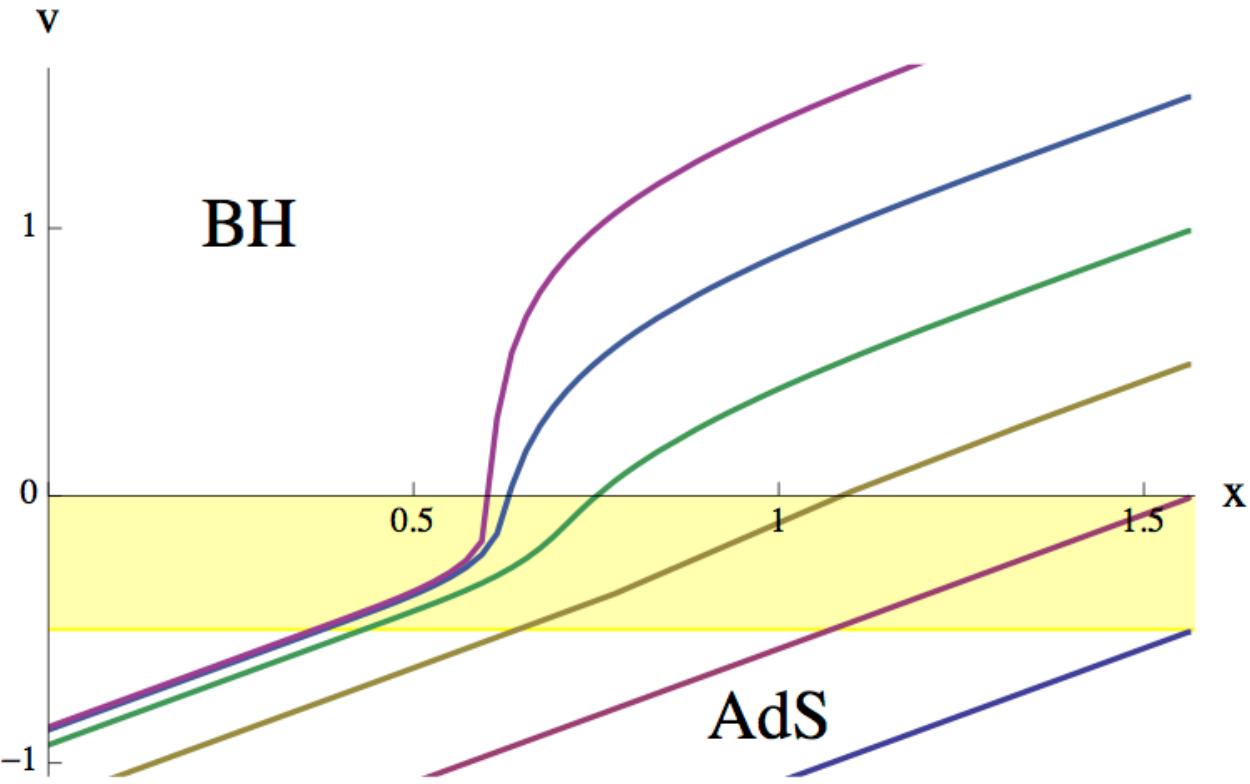} 
\includegraphics[width=7.5cm]{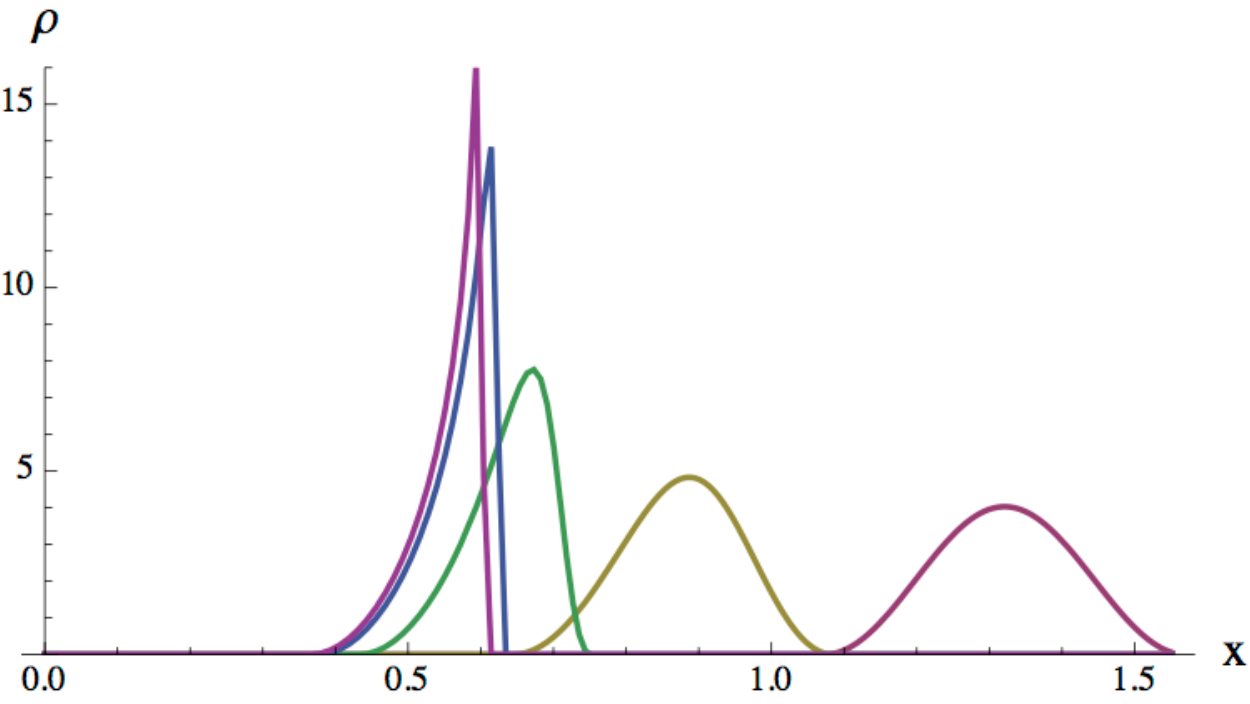} 
\end{center}
\caption{\label{fig:vaidya} Left: intersection a null dust shell with $\Delta=1/2$, signaled in yellow, with the lines of constant $t\!=\!-0.5,0,\dots,2$ in the $(v,x)$ plane. Right: mass distribution function at the same $t$ slices.}
\end{figure}

To gain insight we can resort again to the familiar case of the Vaidya metric
\be
ds^2={1 \over \cos^2 x} \left[ -\left(1 - m(v) {\cos^2 x \over \tan x} \right) d^2v +2 dv dx +\sin^2 x \, d\Omega^2_2 \right]
\label{vaidya}
\ee
with $v$ an Eddington-Finkelstein infalling coordinate. We choose a mass function satisfying
\be
m(v) = \left\{ 
\begin{array}{l l}
0 & v< - \Delta t \, ,
 \\[2mm]
M & v> 0 \, .
\end{array}
\right.
\ee
It represents the building up of an energy density $\epsilon\!=\!M$ in a finite time span $\Delta t$ starting from the CFT vacuum. 
The null dust shell sourcing the metric \eqref{vaidya} has support in the region $v\!\in\![-\Delta t,0]$.
Upon transforming to the Schwarzschild  coordinates $(t,x)$, the shell  exhibits a finite broadness in the radial direction. 
This is shown in Fig.\ref{fig:vaidya}a, where the intersection of a succession of slices of constant $t$ with the location of the shell, highlighted in yellow, determines its radial localization. Its profile at several $t$ slices is plotted in Fig.\ref{fig:vaidya}b.

\begin{figure}[h]
\begin{center}
\includegraphics[width=7.5cm]{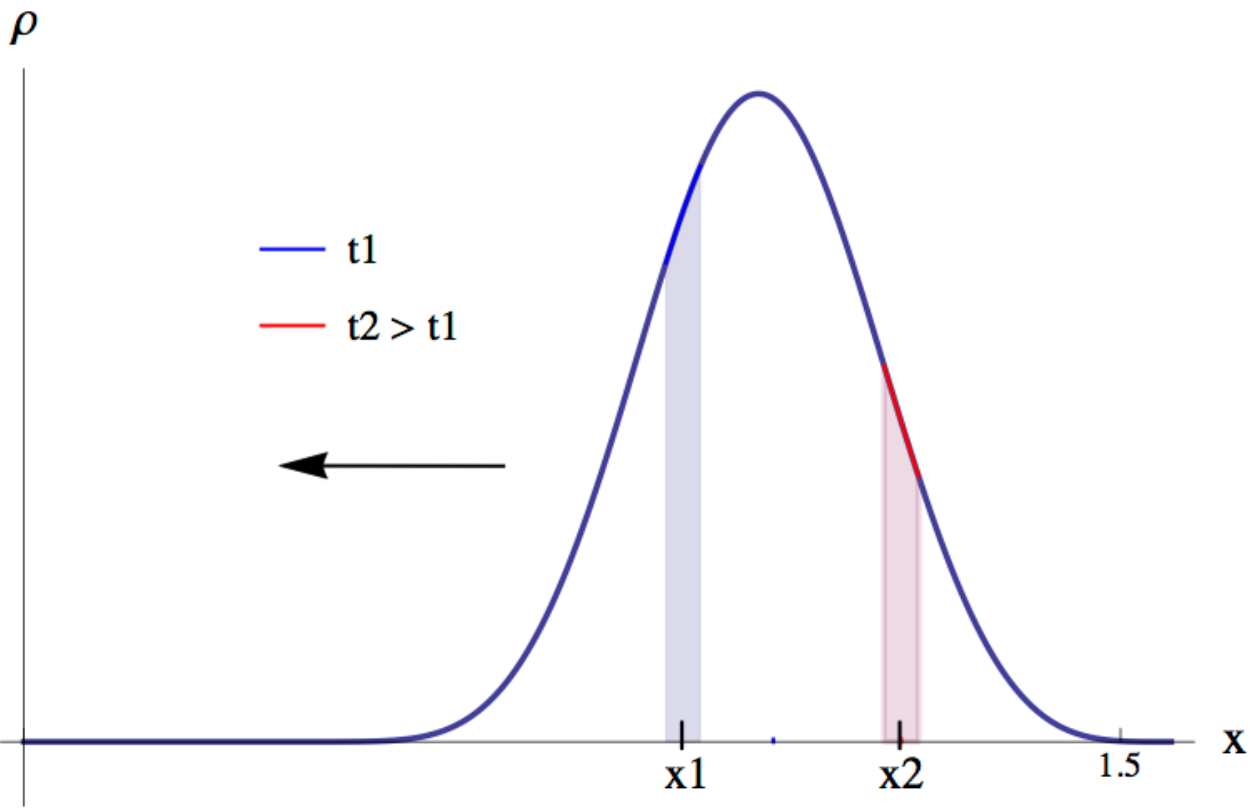} 
\includegraphics[width=7.5cm]{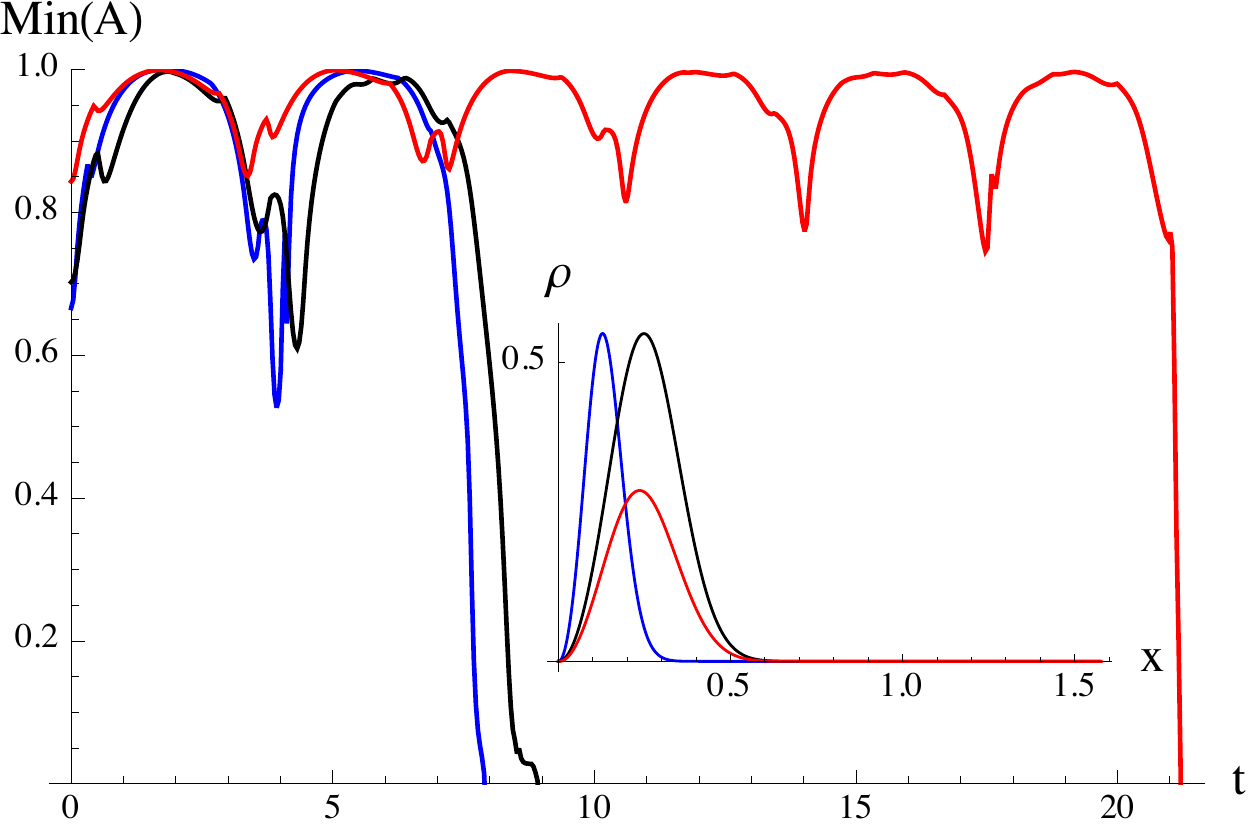} 
\end{center}
\caption{\label{fig:maxent} Left: Excitations sourced during small time intervals around $t_1\!\sim\!x_1\!-\!\pi/2$ and  $t_2\!\sim\!x_2\!-\!\pi/2$.
Right: Evolution of $\min_x A(t,x)$ for several pulses: $\sigma\!=\!0.1$ and $M\!=\!0.035$ (blue), $\sigma\!=\!0.2$ and $M\!=\!0.066$ (black)
and $\sigma\!=\!0.2$ and $M\!=0.035\!$ (red). In the inset we plot the initial mass density distribution for each pulse.}
\end{figure}

The mass distribution function in Fig.\ref{fig:vaidya}b corresponding to $t\!=\!0$
provides the analogue of the initial data we are dealing with in this paper. Its value at a given $x$ reflects the density of excitations created at a prior time  
\be
t \sim x- \pi/2   \leq  0\, .
\ee
The earlier some excitations have been created, the further entangled components are able to fly apart and the deeper its holographic representation reaches in the $x$ direction. We will adopt this point of view in order to interpret the scalar profiles, as sketched in Fig.\ref{fig:maxent}. 
Hence the dual field theory state associated to a broad pulse should describe a configuration with entanglement over many length scales.

The field theory dual to the collapse processes we are considering is in a pure state along the complete evolution (see below) \cite{AbajoArrastia:2010yt}\cite{Takayanagi:2010wp}. One pertinent question is: how much correlation exists between the components $x_1$ and $x_2$ that build up the profile in Fig.\ref{fig:maxent}a? 
Let us recall the evolution of the initial profile \eqref{profileOB}, which is composed of two well localized subpulses.
When the overlap between subpulses is negligible, the gravitational dynamics renders them transparent to each other, see Fig.\ref{fig:int}b in Section 2. This is however not the case for the profile in Fig.\ref{fig:int}a, where the pulses have a small overlap. Both effects
point to the existence of entanglement among the adjacent components of the profile in a way that grows with their proximity. 
This point deserves a deeper investigation.

On general grounds, the time that  a quantum system takes to dephase should depend on the amount of strongly correlated components it contains rather than on the total energy density. 
According to our interpretation, the height in $\rho(0,x)$ provides a qualitative measure of the number of initially strongly correlated excitations.
Hence the time for horizon formation, to be related with the dephasing time in the dual field theory, must be influenced by this value. This is  what we find in  
 Fig.\ref{fig:maxent}b, where we plot the evolution of three different pulses whose initial distributions can be seen in the inset. One of them, blue, has half the broadness of the other two. It coincides with the black one in the  value of the initial amplitude, while with the red one in the total mass. The time of horizon formation is very similar for the pulses with the
 same amplitude (blue and black). However when we compare pulses of the same mass, it is much longer for the broader one (red).\footnote{See \cite{Buchel:2013uba} for a systematic study of the collapse time upon varying the amplitude and the broadness.}

\section{Entanglement entropy oscillations}

All the information necessary to describe the evolution of the dual field theory can be derived from knowledge of the metric and the scalar field.
However, local observables on the quantum field theory only require knowledge about  the asymptotic behavior of the bulk fields. The deep interior geometry can only be accessed through the imprint it leaves on non-local  observables. Following much of the recent literature, we will use to this aim the entanglement entropy. It is defined as the von Neumann entropy of the reduced density matrix for a certain subregion $A$ of the system
\be
S_A= - {\rm Tr}_A (\rho_A \ln \rho_A).
\label{EvN}
\ee
We have already mentioned that the entanglement entropy  has a simple holographic representation in terms of the area of the extremal surface $\gamma_A$ that anchors on the AdS boundary at the boundary of the region $A$ \cite{Ryu:2006bv}\cite{Hubeny:2007xt}. 

For  global AdS$_4$, the boundary  is ${\mathbb R} \times S_2$.  
In the present paper, the regions $A$ we will be interested in are circular caps. Exploiting the symmetry of the problem,  $\gamma_A$ will be a surface of revolution which only depends on a polar angle of the boundary $S_2$.  Hence the task is to find functions $x(\theta)$ and $t(\theta)$ that extremize the area functional subject to the boundary
conditions 
\be
x(\theta_0) = \pi/2  \, , \hspace{5mm}  t(\theta_0)=t_0 \, , \hspace{5mm}  x'(0)=t'(0)=0\, ,
\label{boc}
\ee
where $\theta_0$ is the angular aperture of the cap.
This is a boundary value problem and to solve it we have used a relaxation algorithm \cite{NumRec}.

The entanglement entropy is a UV divergent quantity  and, correspondingly, the area of any surface satisfying \eqref{boc} is also divergent.
The area of the extremal surface anchoring on a cap on empty AdS$_4$ has been obtained explicitly \cite{Johnson:2013dka}
\be
\hbox{Area}(\gamma_A)= 2 \pi \left( \frac{\sin \theta_0}{\epsilon}-1 \right)\, .
\ee
The factor $2\pi$ is due to the rotation symmetry of the configuration.
In order to get a finite result the radial direction has been cut at $x_M\!\lesssim\!\pi/2$. The quantity $\epsilon\!=\!\cot x_M$ has the interpretation of a UV field theory cutoff. Hence the first term in parenthesis gives the area law characteristic of the entanglement entropy \cite{Srednicki:1993im}.
It is then natural to define the finite contribution to the EE by \cite{Albash:2010mv} 
\be
S(t,\theta_0) = \frac{\pi}{2G_4}\left( {\hbox{Area}(\gamma_A) \over 2 \pi} - \frac{\sin \theta_0}{\epsilon}\right)
\ee

The entanglement entropy measured in a pure state satisfies the following important property
\be
S_A=S_{\bar A}
\label{pureEE}
\ee
where $\bar A$ is the region complementary to $A$. The holographic prescription for the calculation of the EE requires that $\gamma_A$ and $A$ be homologous to one another \cite{Fursaev:2006ih}. This leads to the failure of the previous equality in a static black hole background, as it can be expected from the fact that it represents a field theory thermal state. Although the final product of the gravitational collapses we are analyzing is a static black hole, the smoothness conditions   \eqref{exorigin} that we imposed at $x=0$ imply that $\gamma_A$ is homologous to both $A$ and $\bar A$. Thus \eqref{pureEE} is satisfied, reflecting the fact that we are holographically modelling the unitary evolution of a pure state \cite{AbajoArrastia:2010yt}\cite{Takayanagi:2010wp}. The above equality implies that it suffices to study the EE of regions not larger than a hemisphere, namely $\theta_0\!\in\![0,\pi/2]$. 

The portion of spacetime covered by the coordinates $(t,x)$ does not reach behind the apparent horizon. Using a lightlike coordinate $v$ instead of $t$, it has been shown that the extremal surfaces calculating the EE in a Vaidya collapse can  cross both the event and the apparent horizons  \cite{AbajoArrastia:2010yt}. They do so for boundary times and regions whose size is larger than the scale set by the collapsing shell, which is proportional to $M^{-{1\over 3}}$. However we are focussing in scalar configurations for which this scale is larger than the size of the boundary sphere, since these are the only ones that require several bounces  before forming a horizon. Therefore the extremal surfaces we need to calculate will not reach the apparent horizon, and the coordinates $(t,x)$ suffice to describe them.

\begin{figure}[h]
\begin{center}
\includegraphics[width=13cm]{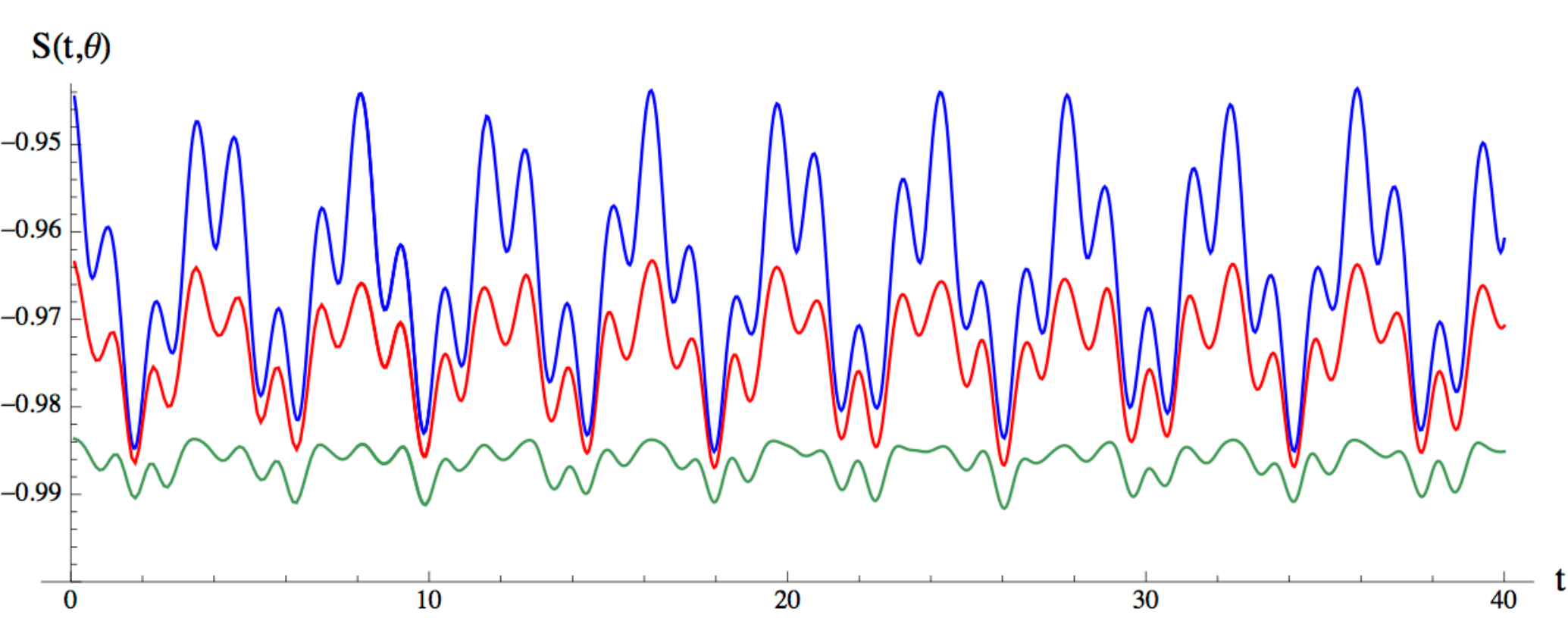} 
\end{center}
\vspace{-3mm}
\caption{\label{fig:oscillations} EE oscillation for an initial profile \eqref{profileO} with $\sigma\!=\!0.4$ and $M\!=\!0.09$. Different colors correspond to caps with angle of aperture   $\theta\!=\!.9,1.2,1.5$.}
\end{figure}

The bouncing geometries induce an oscillating pattern in the entanglement entropy. An example is shown Fig.\ref{fig:oscillations} for a pulse with a regular evolution, namely no horizon seems to emerge, and whose dynamics is somewhat between a quasi-standing wave and a localized pulse. The holographically associated entanglement entropy exhibits oscillations with two clear periodicities, close to $\pi/3$ and $\pi$. These are respectively the periodicities characteristic of the internal dynamics of the pulse and of its radial displacement. 

In the next subsections we will analyze the most relevant features of the holographic entanglement entropy evolution.
Our aim is twofold: we want to learn about the non-equilibrium dynamics of finite size closed systems at strong coupling and, at the same time,  explore the holographic dictionary in dynamical situations. 

\subsection{Early time dynamics}

We shall start our analysis by focusing on the growth of the entanglement entropy as the scalar pulse first falls towards the interior geometry. 
To that purpose we consider narrow initial profiles localized close to the boundary as described by \eqref{profileB}. 

\begin{figure}[h]
\begin{center}
\includegraphics[width=7.2cm]{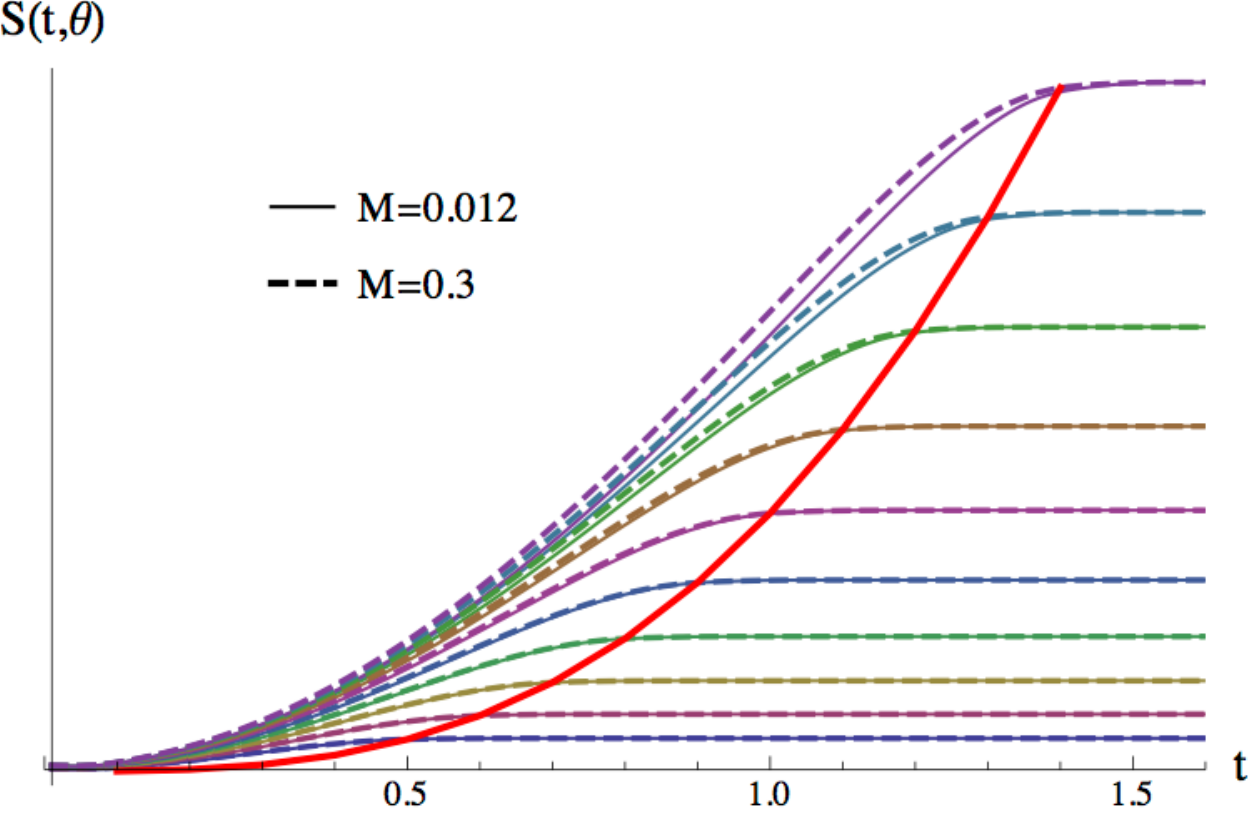} 
\hspace{5mm}
\includegraphics[width=7.2cm]{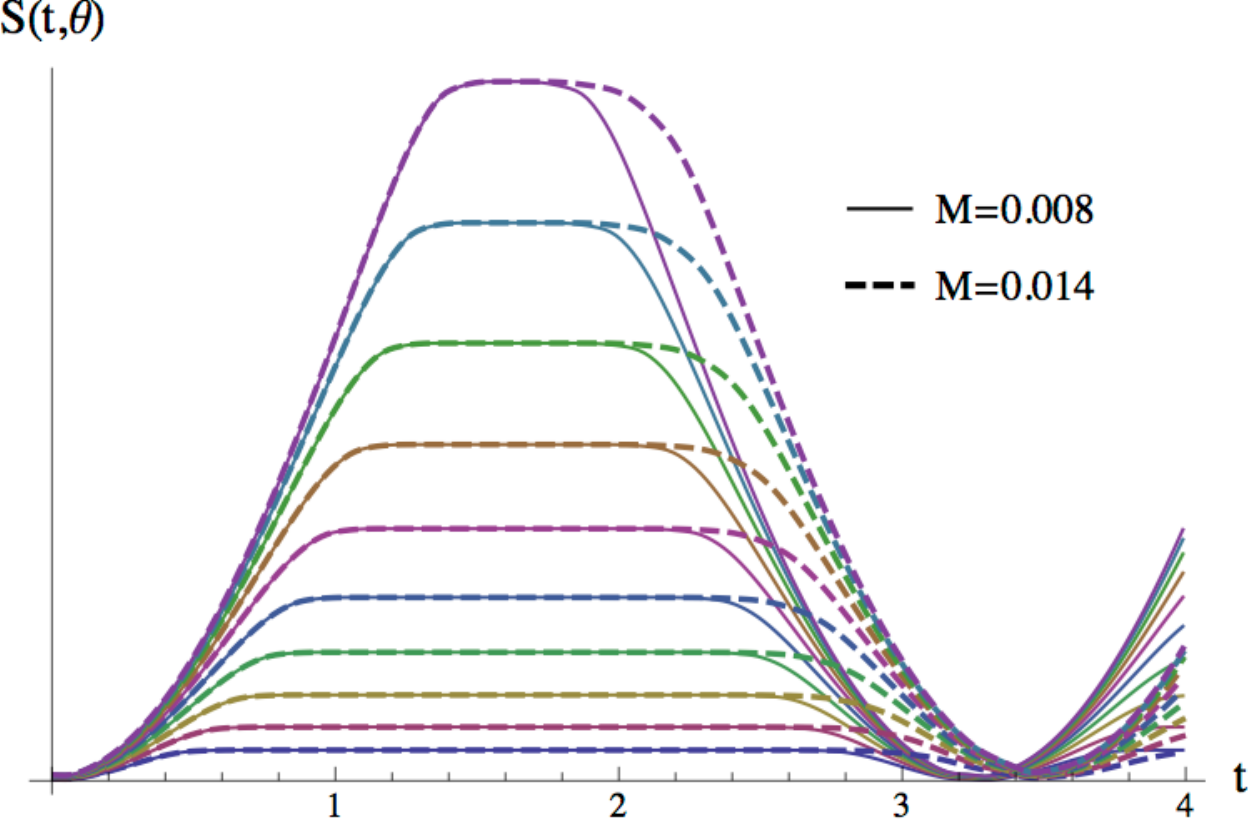}
\end{center}
\caption{\label{fig:early} EE evolution for several pulses with $\sigma\!=\!1/16$. Different colors correspond to caps with $\theta\!=\!.5,.6,..,1.4$. In each graph, the EE values for the lower mass pulse have been rescaled to coincide with those of the larger mass one at their maxima for the sake of comparison.
The red line on the left figure gives, for the pulse with larger mass, the EE at $t=\theta$.}
\end{figure}

In Fig.\ref{fig:early}a we compare two well localized pulses of the same broadness but different amplitudes. One of them gives rise to a black hole of the total mass by direct collapse
 ($M\!=\!0.3$), while the other requires three bounces for the emergence of an apparent horizon ($M\!=\!0.012$). For the sake of comparison, we have rescaled the entanglement entropies of the later case such that they coincide with the former one at their maxima. We find no significant difference between the EE growth to its first maxima for the small mass process and to its final values for the direct collapse one. 
Pursuing this line, in Fig.\ref{fig:early}b we compare a one-bounce ($M\!=\!0.014$) with a many-bounce pulse ($M\!=\!0.008$) using the same rescaling of entropies as before.
In this case there is a perfect match in the growth of the EE for both pulses. Moreover, also the decrease to the subsequent minimum is very similar.
The only important difference is in the time that $S(t,\theta)$ spends at its maximum, which grows with the mass.

These results allow us to sharpen the dual interpretation. The early time dynamics proceeds very much as in non-compact space. Namely, the evolution of the entanglement entropy is qualitatively well described by the propagation model in Fig.\ref{fig:CC}, where the entangled components of the plasma separate at the speed of light. This is illustrated by the red curve in Fig.\ref{fig:early}a. This curve gives the value of the EE at $t\!=\!\theta$, and very approximately signals the moment at which $S(t,\theta)$ saturates to its maxima. Moreover, we have compared the EE growth for narrow shells which form a black hole by direct collapse and Vaidya configurations of approximately the same broadness and mass, observing again no relevant difference.

The effective propagation of entanglement at the speed of light implies that the bulk of entangled excitations have reached their maximal separation on the two-sphere at $t\!\sim\!\pi/2$. However the period needed by the scalar shell to complete a bouncing cycle is always above, although close to $\pi$.  It is practically $\pi$ for pulses of very small mass and increases for more massive ones, as can be seen in Fig.\ref{fig:early}b. 
A dual heuristic picture for this effect could be as follows. Strong interactions might have generated a phase shift on the field theory wavefunction that effectively induces a larger radius for the two-sphere. Since the entanglement entropy depends on the actual size of the region considered, the only natural imprint of the phase shift on the EE would be to prolong the time interval that it keeps at its maximum values. Being an effect due to interaction, it should increase with the energy density of the state, $\epsilon\!=\!M$. This pattern is precisely what we observe in Fig.\ref{fig:early}b.

\subsection{Holographic evolution}

In this subsection we study the evolution of the entanglement entropy based on radially localized pulses.
The collapse of narrow pulses is led by a transference of the energy towards high momentum modes, such that a fraction of the pulse develops a peak sufficiently sharp to become trapped by an emerging horizon. The remaining pulse is swallowed stepwise by a growing horizon, until a final black hole of the total scalar shell mass sets up.

\begin{figure}[h]
\begin{center}
\includegraphics[width=8.5cm]{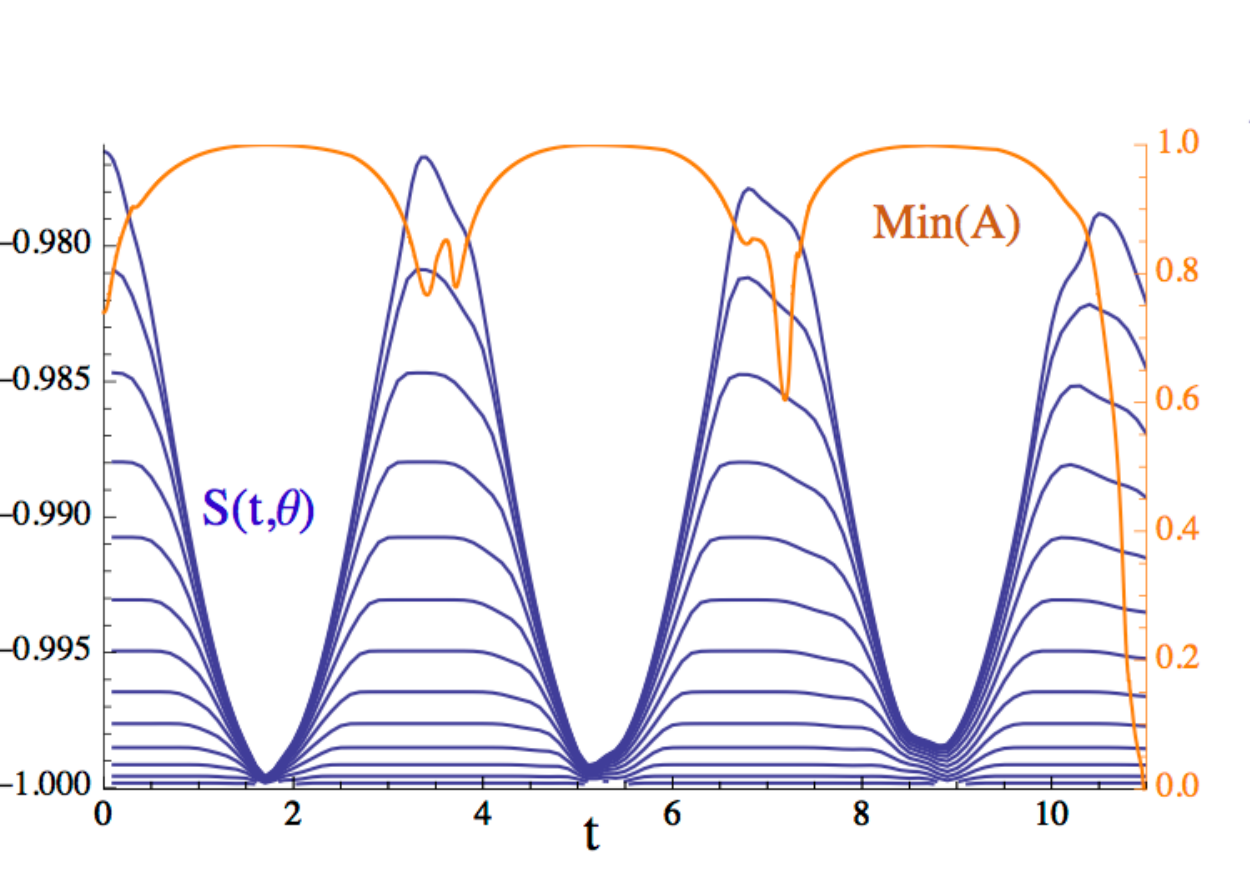} 
\includegraphics[width=7cm]{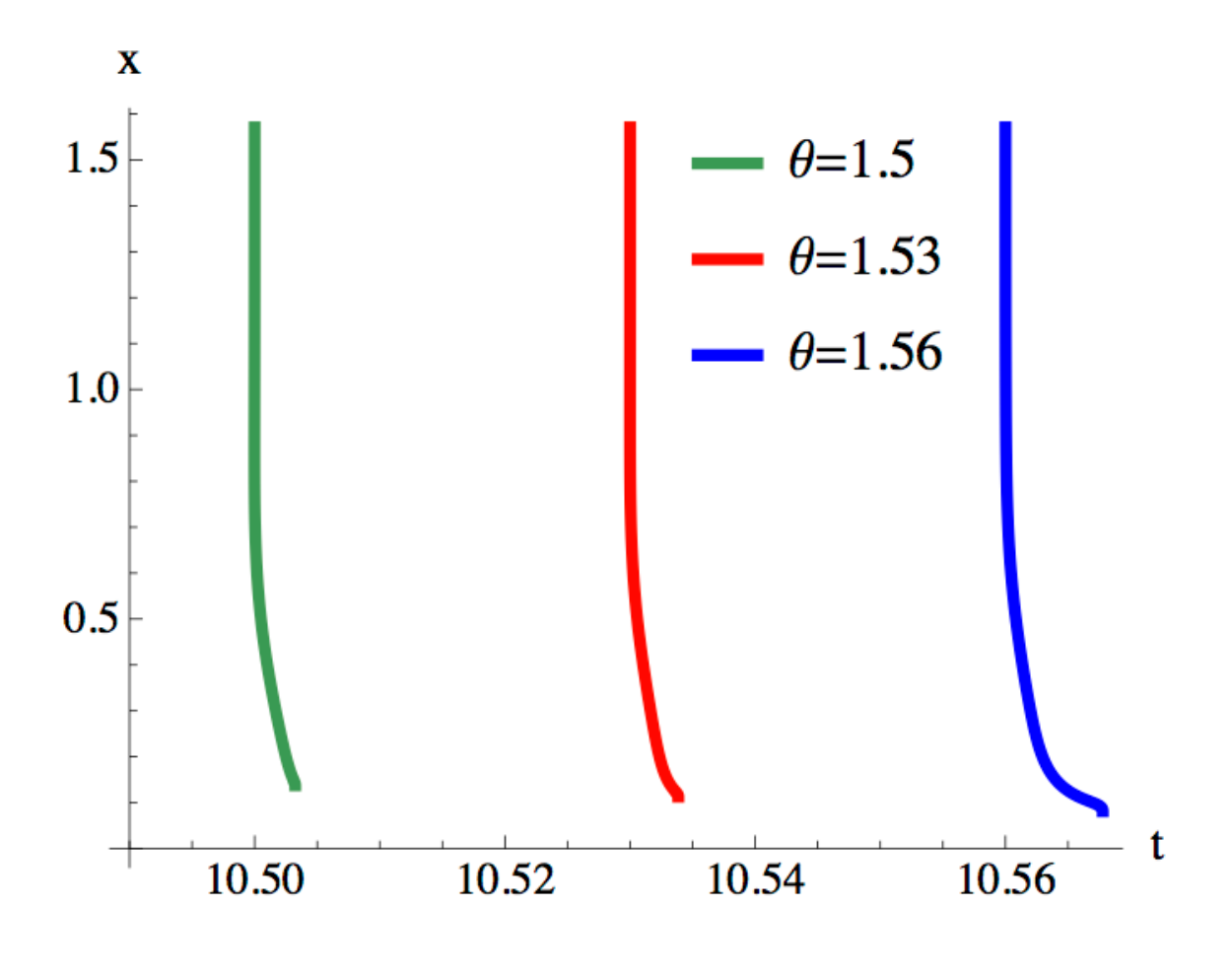} 
\end{center}
\vspace{-5mm}
\caption{\label{fig:maximos_EE} Scalar profile \eqref{profileO} with $\sigma\!=\!0.1$ and $M\!=\!0.012$. Left: Evolution of the
EE for caps  with $\theta\!=\!.5,\dots,1.5$. We have superposed in orange the minimal radial value of $A(t,x)$.
Right: Projection on the $(t,x)$ plane of the surfaces responsible for the EE maxima of large caps in the last bouncing cycle before the horizon forms.}
\end{figure}

Let us analyze first the evolution of the entanglement entropy before a horizon emerges. Remarkably we find that the EE of large regions not only oscillates, but its maxima in each bouncing cycle slightly decrease. We illustrate this effect in Fig.\ref{fig:maximos_EE}a with a narrow pulse which starts close to the origin and requires three bounces to generate a horizon. We showed in Section 2 that when the pulse reaches the origin two opposite effects take place. Namely, together with the sharpening of a fraction of the pulse, the rest tends to increase its radial dispersion, see Fig.\ref{fig:AdSins}a. As a result the extremal surfaces associated to the EE maxima of large boundary regions intersect, at each successive bounce, a growing and more spread fraction of the scalar pulse. 
This causes the   decrease in  area and, hence on $S(t,\theta)$, visible in Fig.\ref{fig:maximos_EE}a.

Although this  is a small effect, we find it relevant. It has been suggested that the entanglement entropy of half the space could provide a definition of coarse grained entropy \cite{Takayanagi:2010wp}. In spite of the oscillations, we might have expected that the maxima of the entanglement entropy monotonically increase along the evolution, and their value still serves as a notion of coarse grained entropy. We have seen that not even this is true in general. 
Recently it has been proposed a different holographic definition of coarse grained entropy \cite{Kelly:2013aja}, related to holographic causal information \cite{Hubeny:2012wa}. It would be very interesting to study its evolution in the collapse processes we are considering.

\vspace{-5mm}

The radial minimum of the metric function $A(t,x)$ is a useful indicator of how far from horizon formation the gravitational system is at a given time slice. The example plotted in Fig.\ref{fig:maximos_EE}a suggests that the maxima of the EE do not necessarily relate to the minima of  $A(t,x)$. Since extremal surfaces do not lie in a constant $t$ slice in our dynamical geometries, we have to analyze what region they explore deep in the bulk before reaching the previous conclusion. For the same example, Fig.\ref{fig:maximos_EE}b shows a projection on the $(t,x)$ plane of the surfaces whose area gives the EE maxima of large caps in the last pre-horizon cycle. They stay indeed well before the time slice where the value of $A$ drops to zero, $t_h\!\sim\!11$. The area of a surface seems to be maximized by a competition between reaching deep in the bulk and keeping outside the traveling shell. After the weak turbulence mechanism has acted on the scalar pulse, the area maximizing configuration arises slightly before $A$ drops to its minimal value. Indeed, the minima along the time evolution of $A$ describe a different situation: the moments at which a well localized peak of the scalar profile is at its closest approach to the origin.
Hence the entanglement entropy turns out not to be precisely correlated with the moment at which a horizon emerges. In particular, it decreases the instants before an apparent horizon first forms.

\begin{figure}[h]
\begin{center}
\includegraphics[width=12cm]{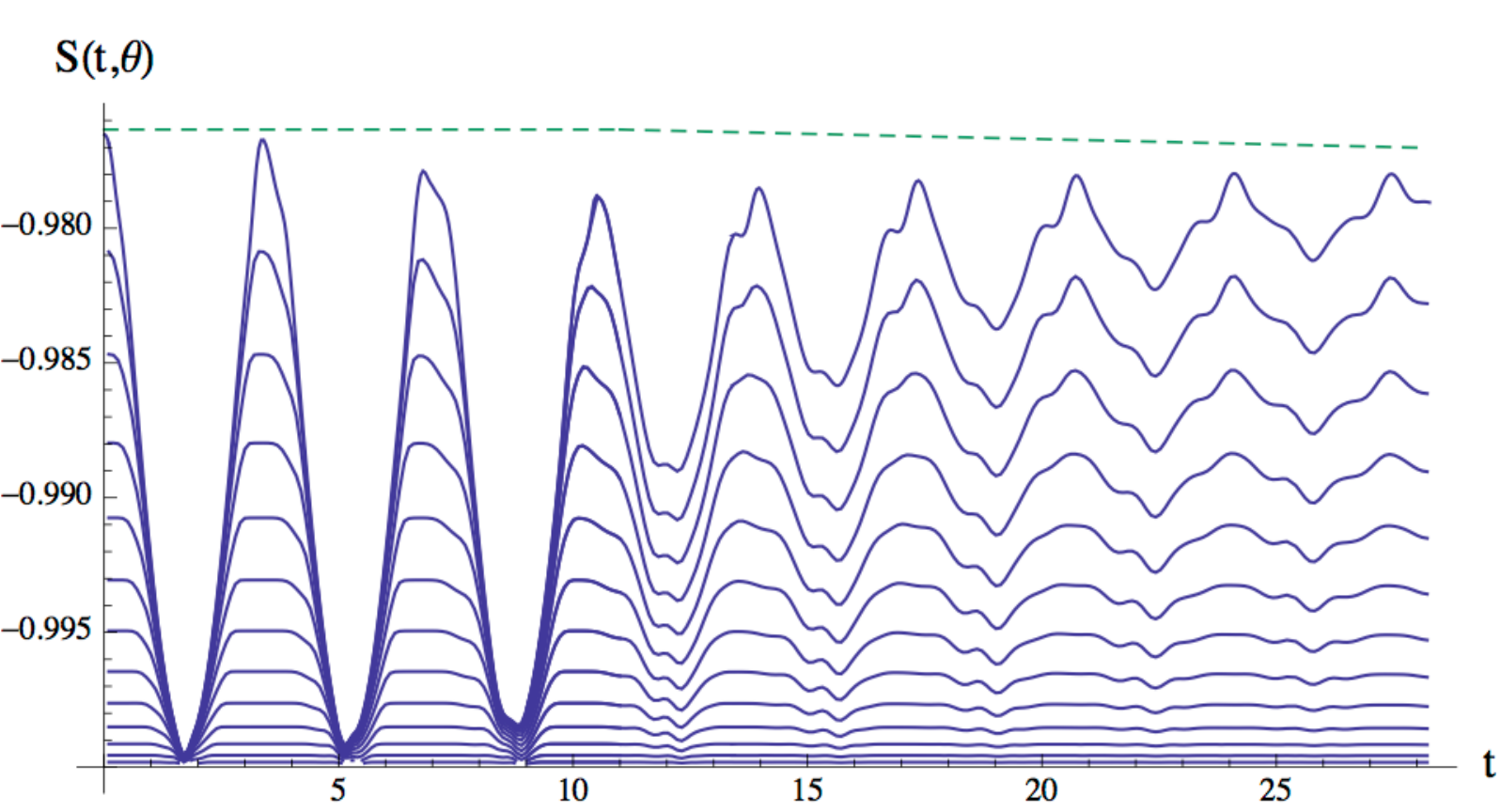} 
\end{center}
\caption{\label{fig:reason} 
Post-horizon evolution of the EE for the same case plotted in Fig.\ref{fig:maximos_EE}. The time for the first collapse  is $t_h \sim 11$, and at around $t = 28$ the horizon radius is   $83\%$ of its final value. The green line gives $S(\theta\!=\!1.5)$ for a static black hole of the total mass, where we did have into account the numerical mass loss along the evolution.}
\end{figure}

It is interesting to describe the behavior of the extremal surfaces with respect to the $t$ slicing.  As long as they do not reach the scalar shell, they live on slices of constant $t$. If an extremal surface intersects a fraction of the falling pulse, the part involved deviates from constant $t$ towards smaller values of the time coordinate. On the contrary, it deviates towards bigger values when it intersects a fraction of the pulse travelling  towards the AdS boundary. 
This is the case in Fig.\ref{fig:maximos_EE}b where the projections in the $(t,x)$ plane show that the extremal surfaces tilt towards larger values of $t$ at their inner portion. Indeed, they reach the part of the scalar profile not trapped by the emerging horizon, and which has started to move away from the origin before the horizon neatly forms.

We have plotted in Fig.\ref{fig:reason} the post-horizon evolution of the entanglement entropy. Using a grid of $7\times 10^4$ points we could complete five  oscillations beyond horizon formation  with a mass loss below $3\%$. 
The oscillations of entanglement entropy  neatly reflect the impact of the horizon,  decreasing their amplitude at a pace correlated with the approach of the horizon to its final value.
The maxima of the EE, whose value dropped along the pre-horizon phase, should rise to the result prescribed by a black hole of the total mass. 
Indeed, we observe that the maxima slowly but monotonically increase along the post-horizon cycles. The green dashed line in Fig.\ref{fig:reason} signals the value that the EE of a $\theta\!=\!1.5$ cap should reach. Its slight decrease just reflects that  we did have into account the small mass loss of the numerically simulation.

The traveling pulse keeps radial localization along the first five post-horizon cycles.\footnote{
Radial localization is manifest in the time span of the oscillations, which is close to $\pi$ before and after a horizon first forms. In spite of this, the radial spread of the profile progressively increases (recall Fig.\ref{fig:AdSpost}a for a similar example). This can be detected in the emergence of a small modulation in the EE with a shorter period, consistent with $\pi/3$.}
Following the argumentation in Section 3, this suggests that while part of the system dephases in correspondence with the appearance of an apparent horizon, part of it still retains quantum coherence. Moreover, a typical separation could be associated to the remaining entangled degrees of freedom, linked to the radial position of the pulse. Thus a pattern emerges in which the system undergoes a stepwise loss of quantum coherence, triggered by the dephasing of a subset of the degrees of freedom. 

Before closing this section, we would like to draw attention to the striking similarity between the oscillations in EE shown in Fig.\ref{fig:reason}, and those of a different, albeit also non-local, operator of the XY quantum spin chain studied in \cite{Igloi2011} (see Fig.1).

\subsection{Behavior across critical points}

A very relevant characteristic in the collapse of narrow pulses is that the fraction of energy in the sharp front which generates the horizon can become arbitrarily small. 
The transition between processes with $n$ and $n\!+\!1$ bounces happens indeed as the energy of the trapped front vanishes. Therefore, it becomes relevant to investigate the behavior of the entanglement entropy across these critical collapses. The question we want to answer is whether the appearance of a horizon, no matter how small, leaves an imprint in the posterior evolution of the entanglement entropy. On line with the results in the previous subsection, the answer we find is negative.

\begin{figure}[h]
\begin{center}
\includegraphics[width=8cm]{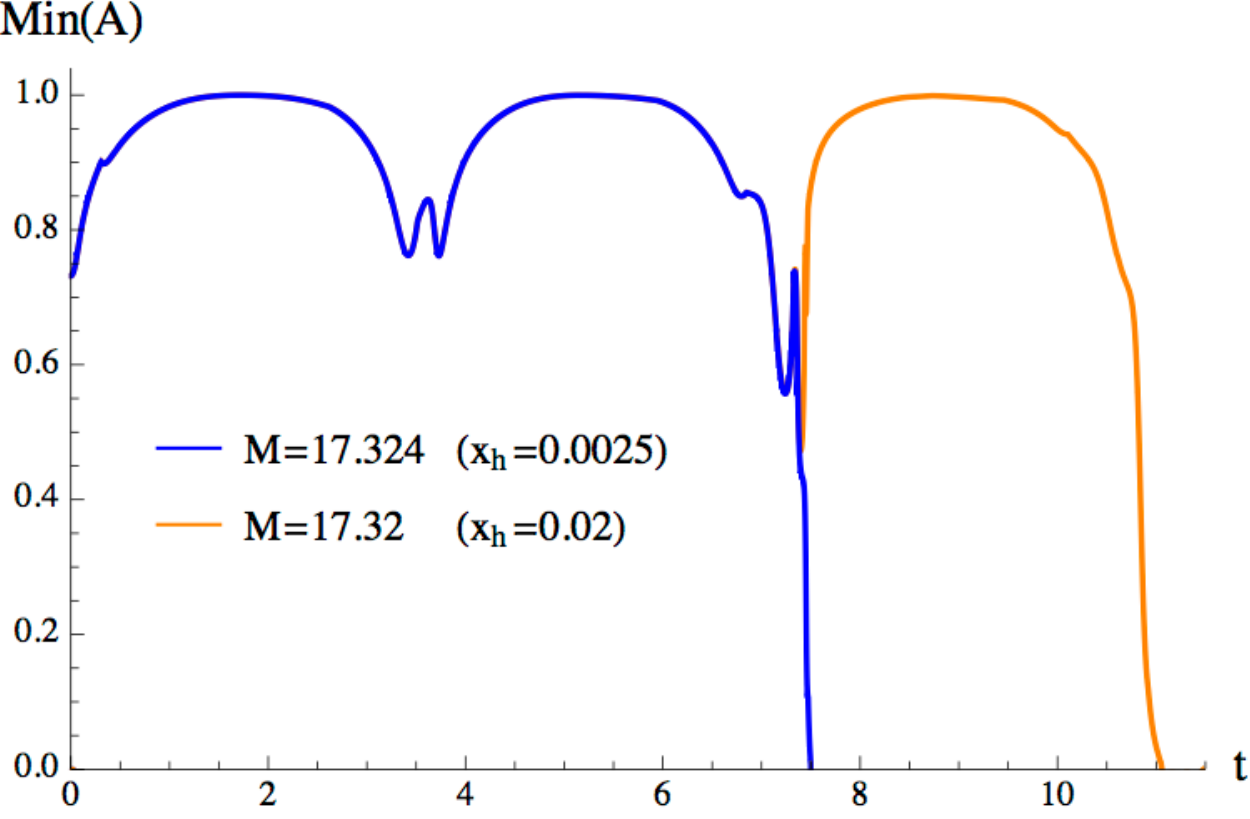} 
\includegraphics[width=8cm]{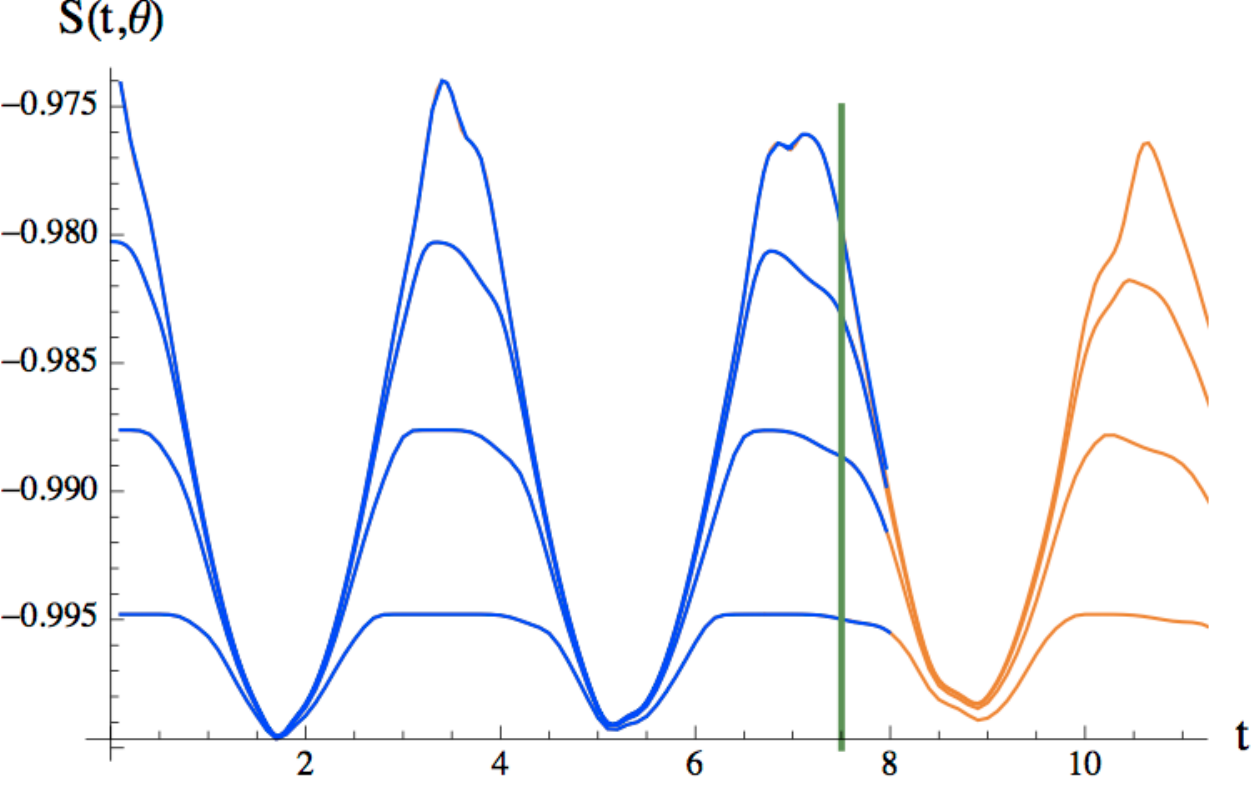} 
\end{center}
\caption{\label{fig:critical} Two pulses with $\sigma\!=\!0.1$ and masses slightly above (blue) and below (orange) critical collapse. Left: minimal radial value of $A(t,x)$. Right: EE evolution for caps with $\theta=0.9,1.2,1.4,1.56$. The green line marks horizon formation time for the above critical pulse.}
\end{figure}

In Fig.\ref{fig:critical}a we plot the evolution of the radial minimum  of $A(t,x)$ for two initial profiles \eqref{profileO} with broadness $\sigma\!=\!0.1$ and masses $M\!=\!17.324$ and $M\!=\!17.32$. The former generates a horizon after two bounces with radius  $x_h\!=\!0.0025$, which is $4.4\%$ of the Schwarzschild radius associated to its total mass.  A trapped horizon emerges for the latter after three bounces with a radius one order of magnitude larger, $x_h\!=\!0.02$, which is $35\%$ of its corresponding Schwarzschild radius. Hence the masses of the two profiles are close to the critical value for the transition between two and three bouncing pre-horizon cycles, being the former slightly above and the latter slightly below. Pursuing the evolution of the profile above critical past the time when the value of $A$ abruptly drops, at $t_h\!\sim\!7.5$, is numerically very demanding. With a grid of $10^5$ points we could only prolong one further time unit while keeping within an acceptable precision. 

As can be seen in Fig.\ref{fig:critical}b, there is no difference between the oscillations of the entanglement entropy in the two cases, both before $t_h$ and shortly afterwards, even for spherical caps very close to a hemisphere.

\subsection{Dependence on the initial state}

\begin{figure}[h]
\begin{center}
\includegraphics[width=13cm]{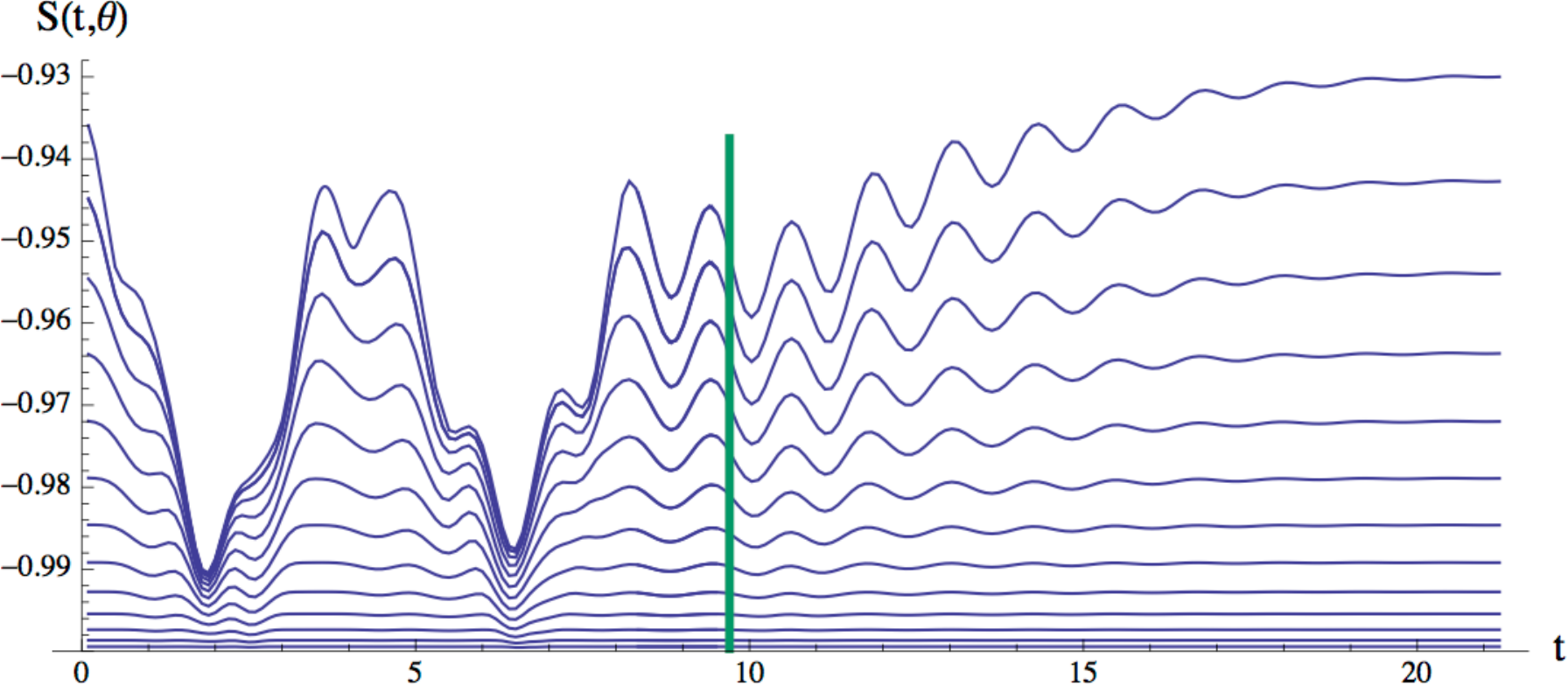} 
\end{center}
\caption{\label{fig:EEmedium} EE evolution to its equilibrium values for caps $\theta\!=\!.5,..,1.5$ derived from an initial scalar profile \eqref{profileO} with $\sigma\!=\! 0.25$ and $M\!=\!0.036$. The time at which ${\rm min}_x A(t,x)\!<\!0.006$ has been signaled in green.
}
\end{figure}

We analyze now how the evolution of the entanglement entropy is influenced by the shape of the scalar profile, which according to our arguments mainly relates to the entanglement configuration of the initial state. We have focussed above on sharply localized pulses. We will consider now the effect of an increasing broadness by taking as example the two pulses studied in Fig.\ref{fig:medium}, of broadness $\sigma=0.25$ and $\sigma=0.6$.

The $\sigma=0.25$ pulse exhibits some degree of radial localization previous to the emergence of an apparent horizon, while its post-horizon dynamics is that of a quasi-standing wave (see Fig.\ref{fig:medium}a).
These two regimes are clearly distinguished by the entanglement entropy, which we show in Fig.\ref{fig:EEmedium}. The green line signals the time at which the radial minimum of $A(t,x)$ drops below $0.006$: $t_h\!\sim\!9.7$. Before $t_h$ the EE oscillates with a period of roughly $4.5$ units (hence $>\pi$), and moreover its value at the local maxima decreases along the evolution. 
On top of these characteristic features  of narrow pulses, the emergence of a shorter modulation can be clearly appreciated.
After $t_h$ only oscillations with a periodicity close to $\pi/3$, proper of a radially delocalized dynamics, are present. They are in one to one correspondence with the minima of $A$ shown in the inset of Fig.\ref{fig:medium}a. The damped nature of the post-horizon evolution is reflected in the decrease of the EE oscillation amplitude. Its maxima monotonically approach the equilibrium value corresponding to a Schwarzshild black hole of the total scalar profile mass. The approach is more efficient in this case than in the narrow pulse of Fig.\ref{fig:reason}, which retains some radial localization along the post-horizon evolution.

\begin{figure}[h]
\begin{center}
\includegraphics[width=8cm]{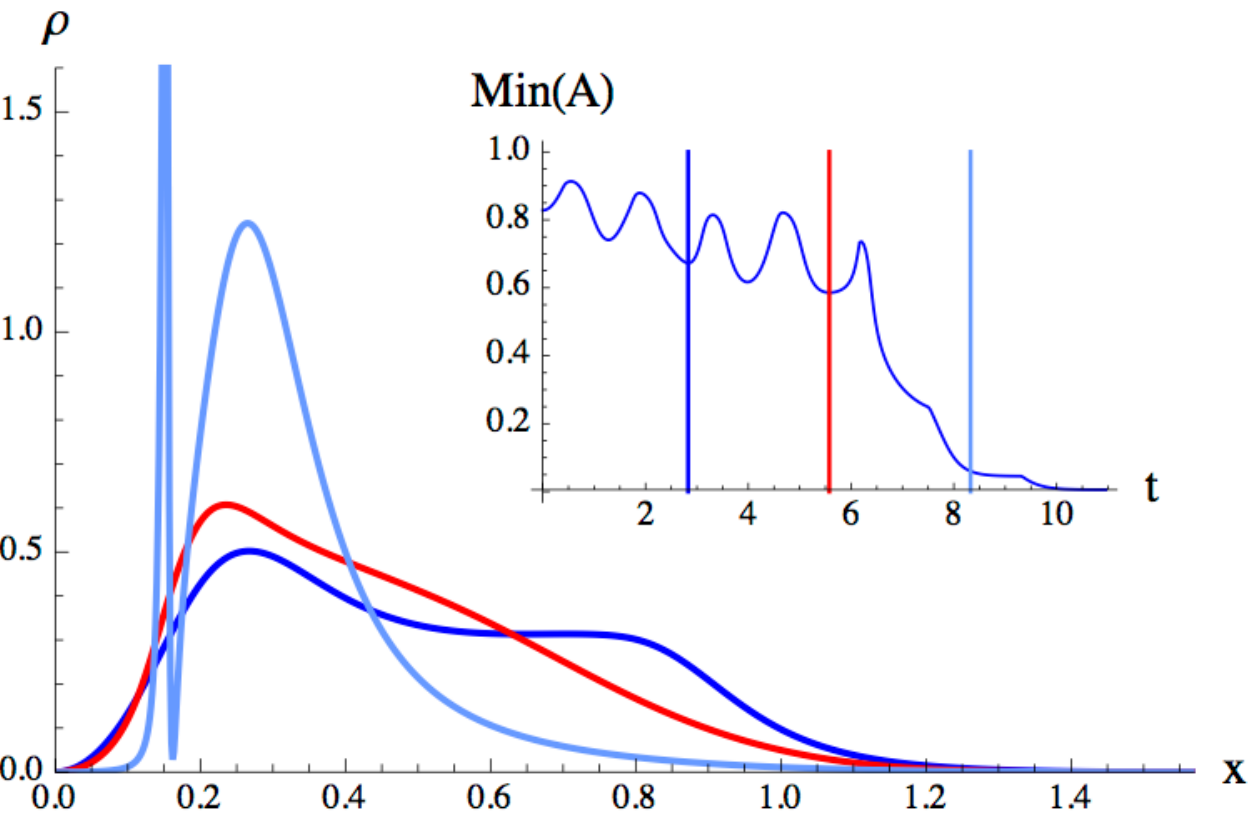} ~
\includegraphics[width=8cm]{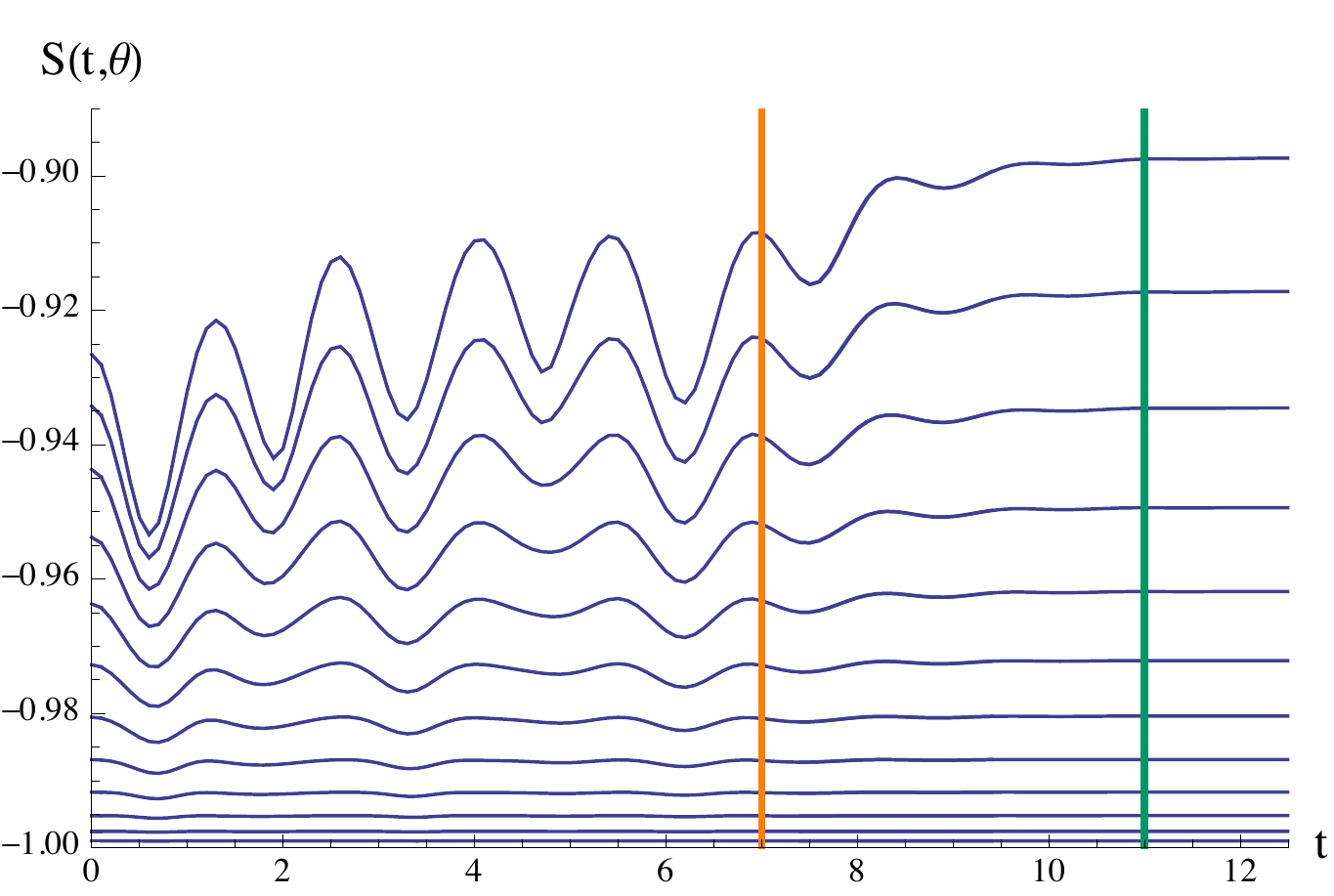} 
\end{center}
\caption{\label{fig:EEbroad} Initial profile \eqref{profileO} with $\sigma\!=\! 0.6$ and $M\!=\!0.1538$. Left: three snapshots in the evolution of the scalar pulse. Right: EE for caps $\theta\!=\!.5,..,1.5$. The time at which the quasi-stationary regime sets in has been signaled in orange and $t_h$ in green.
}
\end{figure}

The $\sigma\!=\!0.6$ pulse in Fig.\ref{fig:medium}b is radially delocalized along its complete evolution. Its mass is $40\%$ the threshold mass for large AdS$_4$ black holes. Hence it develops quite massive subpulses, that delay in climbing their gravitational potential. 
This favors the recombination of subpulses into a single very broad peak, as can be observed in the first two snapshots of Fig.\ref{fig:EEbroad}a. At some point this dynamics gives way to the establishment of a strongly damped oscillation. It happens at $t\!\sim\!7$  for the example in Fig.\ref{fig:EEbroad}a. The third snapshot of the mass distribution profile corresponds to this regime. The minimum radial value of $A(t,x)$ drops below $0.006$ at $t_h\!\sim\!11$, when practically the complete scalar pulse has been trapped by the emergent horizon. 

The evolution of the entanglement entropy for this pulse is shown in Fig.\ref{fig:EEbroad}b. The EE maxima do not decrease along the evolution. The amplitude of the EE oscillations starts decreasing once the damped quasi-stationary regime sets in. At $t_h$ the EE has practically reached its equilibrium values. Throughout the complete evolution the oscillations have  invariably a period $\gtrsim\!\pi/3$.

Finding a qualitative field theory explanation for this period seems difficult. Its gravitational origin is in the internal dynamics of the scalar profile rather than on the radial propagation.
From the field theory point of view, this hints towards having its root in the strong coupling dynamics of the out of equilibrium plasma.

We have argued that the collapse of narrow pulses describes a stepwise relaxation process, triggered by the dephasing of a subsystem. The evolution of broad pulses suggests a quite different mechanism. The  dynamics leading to the strongly damped oscillations involves in this case the whole system. We proposed in Section 3 that broad pulses correspond configurations with entangled excitations over many length scales. Hence we conclude that, in such situation, dephasing triggered by a subsystem is disfavored. 
Moreover, when the mass of a broad pulse drops below a certain threshold no horizon appears to be formed \cite{Buchel:2013uba}. This would indicate that reaching a stationary state is harder for states of small energy when this involves the complete system, than when it affects only a subsystem. For the former case, and under the assumption of a spherically symmetric collapse, below a certain energy density the dual field theory system seems to never lose quantum coherence.

\section{Conclusions}

The out of equilibrium dynamics of closed quantum systems is a subject attracting an increasing attention thanks to the experimental control over systems involving cold atoms and boson condensates. It turns out that a fast approach to ergodic behavior is not always realized. Some systems exhibit a long lived far from thermal quasi-stationary stage, known as prethermalization plateau \cite{Gring.et.al.2012}\cite{Trotzky.et.al.2012}, and only achieve thermal equilibrium on a longer time scale.   
Besides, the dephasing time in finite size setups can be sufficiently large to allow for several partial reconstructions of the initial state and lead to an oscillating behavior of some observables.  An outstanding experimental example was realized in \cite{Kinoshita2006}. The aim of the present paper was to start the study of this rich phenomenology by means of holographic techniques.

Following previous works, we have considered the evolution of a massless scalar field coupled to Einstein gravity with a negative cosmological constant in four dimensions. 
We focused on global coordinates such that the dual theory lives on a two-sphere, and restrict to spherically symmetric bulk configurations.
We gave arguments which support that, for small enough amplitudes, this simple setup is able to model relaxation processes with large dephasing times. 
Contrary it is not suited to study the above mentioned prethermalization stage. Such, far from thermal, intermediary regime can typically be understood as an equilibrium state in an integrable theory closely related to the system under study \cite{Jaynes1957}\cite{Kollar2011}\cite{Smith2012}. However a self gravitating real scalar configuration only carries one conserved charge, the mass. It would be very interesting to address this problem in the context of higher spin theories in AdS. 

Our main proposal is that gravitational collapse processes requiring several bounces off the AdS boundary to generate a horizon \cite{Bizon:2011gg}\cite{Buchel:2012uh},
map holographically to field theory evolutions exhibiting large dephasing times.
The radial position of the collapsing scalar shell would encode the typical separation of entangled components in the dual out of equilibrium plasma. When the pulse is close to the boundary, entanglement is concentrated among nearby excitations. Its fall towards the interior geometry is the picture dual to entangled excitations flying apart at the speed of light, as in the holographic model for a local quench of \cite{Nozaki:2013wia}. The travel of the shell back to the boundary will model entangled components approaching again on the boundary sphere. 
This interpretation is on line with recent works directly linking the entanglement pattern of a field theory with a higher dimensional geometry \cite{VanRaamsdonk:2009ar}. In particular, entanglement entropy is used in \cite{Nozaki:2013vta}\cite{Lashkari:2013koa} as a suitable observable to translate Einstein equations into field theory language. Going further, in \cite{Swingle:2009bg}\cite{Nozaki:2012zj} the authors propose to build up a geometry from a quantum field theory state using the Multiscale Entanglement Renormalization Ansatz \cite{Vidal} (MERA) and its continuum version (cMERA) \cite{Haegeman:2011uy}. 

We have also resorted  to the entanglement entropy as a probe of the field theory out of equilibrium dynamics. The entanglement entropy evolution broadly  follows the one for the metric. The oscillatory pattern that we have found is a reflection of this fact. For narrow scalar profiles, the periodicity of the EE oscillations is always close to  $\pi$. This is indeed the time that two particles, travelling at the speed of light, take to reunite again along an equator of the boundary sphere, and provides the main support for the advocated picture. Consistently the minima of the EE happen when the scalar shell bounces at the boundary, while its maxima are reached when the shell is close to the origin. 

The entanglement entropy is not only able to detect propagation, but also interaction effects. They are related to the internal dynamics of the shell. A main characteristic of this dynamics is weak turbulence, that governs the appearance of a horizon for profiles of small mass.
We have stressed that, besides the generation of a sufficiently sharp front for collapse, there is an accompanying effect of radial dispersion in the scalar profile leftover by the emerging horizon. Interestingly, this flow of energy towards opposite scales in the bulk, is reflected in a decrease of the entanglement entropy maxima along the pre-horizon evolution. 

In order to get a complete picture of the field theory relaxation process, we continued the gravitational evolution until a static black hole is almost established. 
For narrow initial pulses, the fraction of the scalar profile escaping from the emergent horizon exhibits radial localization along several further bouncing cycles.
According to our interpretation this suggests that while part of the system has dephased, a subset of degrees of freedom keeps quantum coherence. Moreover, a typical separation can be associated to the remaining entangled components, which still evolves with a period close to $\pi$. This stepwise pattern of relaxation is an important result of our analysis. 

The amplitude of the entanglement entropy oscillations neatly reflects the formation of a horizon, decreasing as the trapped mass increases. In spite of this, the emergence of a horizon, understood as the region in the geometry where the metric function $A(t,x)$ first drops to an almost vanishing value, does not leave a sharp imprint on the entanglement entropy. Neither do the regions of its stepwise growth. Along the post-horizon evolution the entanglement entropy rather exhibits damped but smooth oscillations.

The holographic set up we have considered has a number of limitations\footnote{We thank M. Rangamani for a discussion on this issue.}. Our simple bulk theory is not motivated by a consistent truncation of a string derived supergravity. 
On the other hand, if small AdS black holes are embedded in ten-dimensional supergravities, they exhibit a Gregory-Laflamme instability below a certain mass threshold \cite{Hubeny:2002xn}. In addition, we have focussed on a very restricted set of initial data, those with spherical symmetry.
In spite of this shortcomings, we expect that the qualitative picture we have put forward persists when the simple bulk theory is generalized, or when spherical symmetry is dropped. This actually amounts to assume that a generic matter configuration with pressure and sufficiently small mass, will not generate a horizon by direct collapse.  In an asymptotically global AdS space, provided the weak turbulence mechanism is efficient for a wide range of initial data, an apparent horizon should emerge after enough reflections off the boundary. That would trigger the approach to a final static black hole state, or equivalently, to an ergodic regime in the dual strongly coupled field theory. 

The present work points up in several directions worth exploring further. Certainly, a more accurate numerical simulation of the 
post-horizon evolution, including the use of horizon penetrating coordinates is in order. The transition to stable bouncing solutions is a very interesting question to investigate, presumably representing a transition to systems that never thermalize.

 Also very interesting is the extension of the present work to other types of operators in the quantum theory, dual to massive scalars, and or gauge fields.
Finally, the case of AdS$_3$ is very important \cite{Bizon:2013xha}, given the existence of explicit results in the dual conformal field theory that could help improve the dictionary. Numerically this case is difficult to tame, but we plan to report on it soon.

\section*{Acknowledgements}
Many special thanks go to Sascha Husa, how helped us in developing and testing the codes for the non-linear evolution of the equations of motion. Also we want to thank Jos\'e L.F. Barb\'on, Alex Buchel, Vitor Cardoso, Oscar Dias, Roberto Emparan, Veronika Hubeny, Javier Molina-Vilaplana, \'Angel Paredes,  Bel\'en Paredes, Mukund Rangamani, Subir Sachdev and Germ\'an Sierra.  The work of E.L. has been supported by the Spanish grant FPA2012-32828, Consolider-CPAN (CSD2007-00042), and SEV-2012-0249 of the Centro de Excelencia Severo Ochoa Programme. The work of J.M.  is supported in part by the Spanish grant  FPA2011-22594,  by Xunta de Galicia (GRC2013-024), by the  Consolider-CPAN (CSD2007-00042), and by FEDER. He also wishes to express his gratitude to the Kavli IPMU at  Kashiwa, Tokyo, were part of this project was carried out, thanks to a visit funded  by the 7th Framework European Programme Unify [FP7-People-2010-IRSES- grant agr. No 269217]. J.A-A. is supported by an IFT-Severo Ochoa contract SEV-2012-0249. E.daS. is supported by an IFT-Severo Ochoa contract SEV-2012-0249 and by the Spanish grant FPA2012-32828. A.S. is supported by the European Research Council grant HotLHC ERC-2011-StG-279579 and by Xunta de Galicia (Conselleria de Educaci\'on). 

\end{document}